\documentclass[twocolumn,english,prl]{revtex4-1}
\usepackage{color}
\usepackage{graphicx}
\usepackage[T1]{fontenc}
\usepackage{float}
\usepackage{textcomp}
\usepackage{amsmath}
\usepackage{amssymb}
\usepackage{graphicx}
\usepackage{esint}
\usepackage{natbib}
\usepackage{color}

\begin{document}

\title{Spin and valley control in single and double electrostatic silicene quantum dots}

%% Notice placement of commas and superscripts and use of &
%% in the author list

\author{Bart{\l}omiej Szafran }

\affiliation{AGH University of Science and Technology, Faculty of Physics and
Applied Computer Science,\\
 al. Mickiewicza 30, 30-059 Krak\'ow, Poland}

\author{Dariusz \.Zebrowski }

\affiliation{AGH University of Science and Technology, Faculty of Physics and
Applied Computer Science,\\
 al. Mickiewicza 30, 30-059 Krak\'ow, Poland}

\begin{abstract}
We study quantum dots defined electrostatically within silicene.
%using an atomistic tight-binding approach including the time evolution 
%of the confined states and the electron-electron correlation in the two-electron system. 
We determine the spin-valley structure of confined single- and two-electron systems,
 and quantify the effects of the  intervalley scattering by the electron-electron interaction potential and  the crystal edge. The double quantum dots are discussed in the context of the spatial symmetry of the extended
orbitals. We determine the charge, spin and valley transitions times induced by alternate electric fields. We show that the valley transition times can be changed within several orders of magnitude by the depth of the confinement potential. 
Also, the spin transition rates can be enhanced by orders of magnitude by the coupling
of the bonding and antibonding orbitals mediated by the Rashba spin-orbit interaction. 
\end{abstract}

\maketitle

\section{Introduction}

Quantum dots with electrostatic confinement defined in III-V materials 
are used for studies of the confined spin control \cite{eqq,fabian}. 
In graphene the spin coherence and relaxation times \cite{vo,weihan,mass,stano}
are long which should make this material  attractive  \cite{gb} 
for applications of confined spins in  quantum computing \cite{divi}.
However, the weakness of the spin-orbit interaction in graphene \cite{laird} excludes
the spin manipulation by electric fields \cite{edsr,extreme,stroer,nadji}. 
On the other hand, the graphene offers the valley  \cite{grafen,ry}  instead of spin \cite{wulu,pal,recher} for information processing.
Nevertheless, the lack of the energy gap makes the purely confinement in pristine graphene
excluded \cite{klein}. 
The gap can be opened in bilayer graphene \cite{review,gap1,gap2,gap3,gap5,gap6} which solves the problem for electrostatic  confinement  \cite{biqd0,biqd1,biqd2,biqd3,biqd4,biqd5,biqd6}. However, the 
manipulation of the spin is still hampered by the weakness
of the spin-orbit coupling  \cite{SoBilFabian}.

An alternative material,
in which both the spin and the valley \cite{Pan14} degrees of freedom 
could be controled by electric fields, is the silicene \cite{silitmdc,chow}. 
As in bilayer graphene, the perpendicular electric field opens the energy gap 
\cite{ni,Drummond12} which allows for electrostatic confinement. Silicene is characterized by 
relatively strong intrinsic spin-orbit interaction 
 \cite{Liu11,Liu,Ezawa}
with the coupling constant which by three orders of magnitude \cite{chow} exceeds the one for graphene. The fact inspires proposals for spin-active transport devices \cite{Xu12,Rachel14,Zutic04,miso15,rzeszot,Tsai13} defined in silicene. 

%As compared to graphene, fabrication of silicene for operating devices
% is more difficult. 
Silicene was first grown on metals \cite{Vogt12,Aufray10,Feng12,siedem,dziewiec} 
and studies of the  electronic properties require non-metallic substrates. Growth of silicene on   AlN \cite{aln}
and   transition metal dichalcogenides (TMDCs) \cite{silitmdc,tmdc1,tmdc2}
was theoretically studied. A succesful fabrication of a
 room-temperature field effect transistor was reported \cite{Tao15} for silicene 
on Al$_2$O$_3$. The Al$_2$O$_3$ substrate preserves the electron structure of free-standing silicene near the Dirac points \cite{al2o3}. Studies of silicene islands 
grown on graphite by van der Waals heteroepitaxy are also performed \cite{acs1,acs2}.

In this work we determine the  spin- and valley- structure of states electrostatically confined  within silicene and consider manipulation of the spin and valley degrees
of freedom using alternate electric fields of microwave or deep infrared frequency. 
The alternate electric fields were previously applied for
 states confined in carbon nanotubes \cite{laird,edsr4,edsr5}.
The experiments \cite{edsr4,edsr5} resolve the spin and valley transitions 
using double quantum dots.
Here, we consider one excess electron and an electron pair in single and double dots. We use the atomistic tight-binding approach that naturally accounts for the valley mixing effects of the crystal edge \cite{zarenia} and the
intervalley scattering due to the short-range component of the Coulomb interaction  
\cite{umklapp,mayrhoer,rontani,pecker}.

Previously, silicene flakes without the electrostatic confinement potential were  studied  \cite{kiku,romera,abdelsalam,szafran18}.
A type-I electrostatic quantum dot, that supports
localization of conduction and valence band states was also discussed \cite{szafranre}. Here, we present a simpler set-up for electrostatic type-II quantum dots -- that supports confinement
of excess electrons in the conduction band only.  We determine the intervalley
scattering effects for the two-electron spectrum, and the tunnel coupling effects
for the spatial symmetry of the wave functions in double quantum dots.
We find that the spectrum for the double dots can be described in terms
of separation of the spin-valley and spatial degrees of freedom.
We demonstrate that both spin and valley transitions can be driven by AC electric
fields and that the valley transition rates can be controlled in a large range by 
the tunable coupling to the edges of the flake. We also indicate 
that avoided crossings open by Rashba interaction between the bonding and antibonding
states in double dots make the spin transition  as fast as the spin conserving transitions.

\section{Theory}

\subsection{Single-electron Hamiltonian}
We determine the single-electron eigenstates for an atomistic tight-binding Hamiltonian \cite{Liu}
which in the absence of the external magnetic field reads
\begin{eqnarray}
H_0&=&-t\sum_{\langle k,l\rangle \alpha }  c_{k\alpha}^\dagger c_{l\alpha} +it_2 \sum_{\langle \langle k,l\rangle \rangle \alpha, \beta } \nu_{kl} c^\dagger_{k\alpha} \sigma^{z}_{\alpha\beta}c_{l\beta} \nonumber \\&& -it_1 \sum_{\langle \langle  k,l \rangle \rangle \alpha,\beta } \mu_{kl} c^\dagger _{k\alpha}\left(\vec{\sigma}\times\vec{d}_{kl} \right)^z_{\alpha\beta} c_{l\beta} \nonumber 
\\ && +it_3 \sum_{\langle k,l \rangle,\alpha,\beta} F_z({\bf r}_k)   c_{k\alpha}^\dagger \left(\vec{\sigma}\times\vec{d}_{kl} \right)^z_{\alpha\beta}  c_{l\beta}  \nonumber 
\\ && +\sum_{k,\alpha} V_k c^\dagger_{k\alpha}c_{k\alpha},  \label{hb0}
\end{eqnarray}
with summations over the nearest-neighbor ($\langle k,l\rangle $), the  next-nearest-neighbor ($\langle\langle k,l\rangle\rangle $) ions,
  and the spins ($\alpha$,$\beta$) of electrons localized at  Si   $3p_z$ orbitals.
We use  $t=1.6$ eV \cite{Liu,Ezawa,chow} for the nearest-neighbor hopping. 
The second term of Hamiltonian   (\ref{hb0}) is the intrinsic spin-orbit interaction
 \cite{Liu,Ezawa,km,chow} -- the dominant spin-orbit coupling term for silicene.
The intrinsic spin-orbit parameter is $t_2=0.75$ meV \cite{Liu,Ezawa},
and  $\nu_{kl}=\pm 1$. The positive (negative) sign of $\nu_{kl}$ is set for   
 the counterclockwise (clockwise) next-nearest neighbor hopping via the common neighbor ion.
The second line in Eq. (\ref{hb0}) introduces the built-in Rashba spin-orbit interaction 
which results from the presence of a perpendicular electric field component for the buckled lattice, 	
with $t_1=\frac{7}{15} $ meV \cite{Liu,Ezawa} and  ${\bf d}_{kl}=\frac{{\bf r}_l-{\bf r_k}}{|{\bf r}_l-{\bf r_k}|}$, 
where ${\bf r_k}=(x_k,y_k,z_k)$ indicates the position of the $k$-th ion,  $\mu_{kl}=\pm 1$, with plus for sublattice A 
and minus for sublattice B. The superscript $z$ above the parentheses 
in the second and third line of Eq. (\ref{hb0}) stands for the $z$-component of the vector operator
defined by the cross product.
 The third line of Eq. (\ref{hb0}) introduces
the extrinsic Rashba interaction due to the external electric field perpendicular to the silicene plane, with $t_3=0.589 \times 10^{-3}$ \AA, which for
$F_z=17$ meV/\AA\; gives $t_3F_z=10$ $\mu$eV \cite{Ezawa}. 

\subsection{Electrostatic quantum dot potential}
\begin{figure}
\begin{tabular}{ll}
(a) \includegraphics[width=0.4\columnwidth]{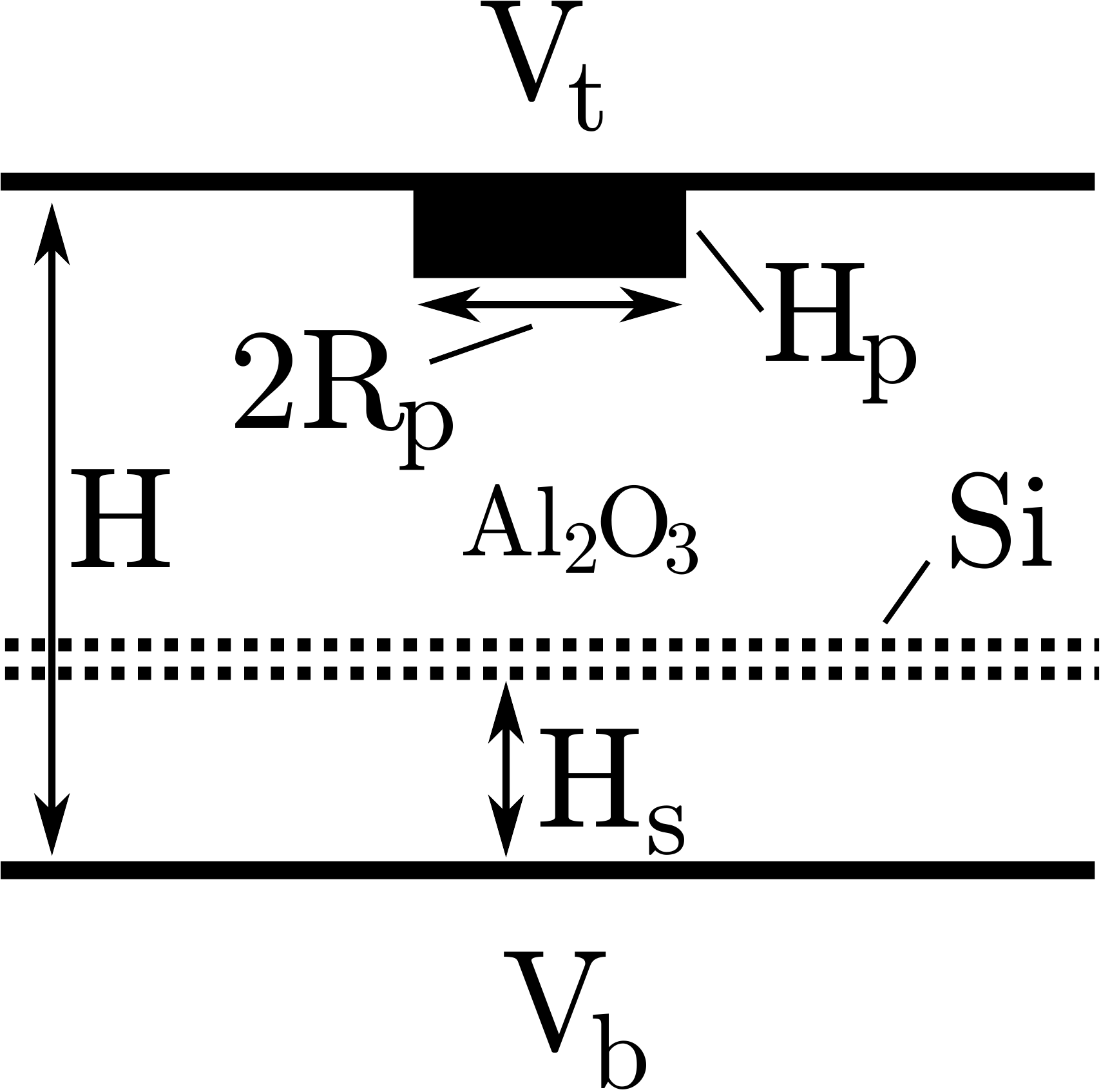} &
(b) \includegraphics[width=0.4\columnwidth]{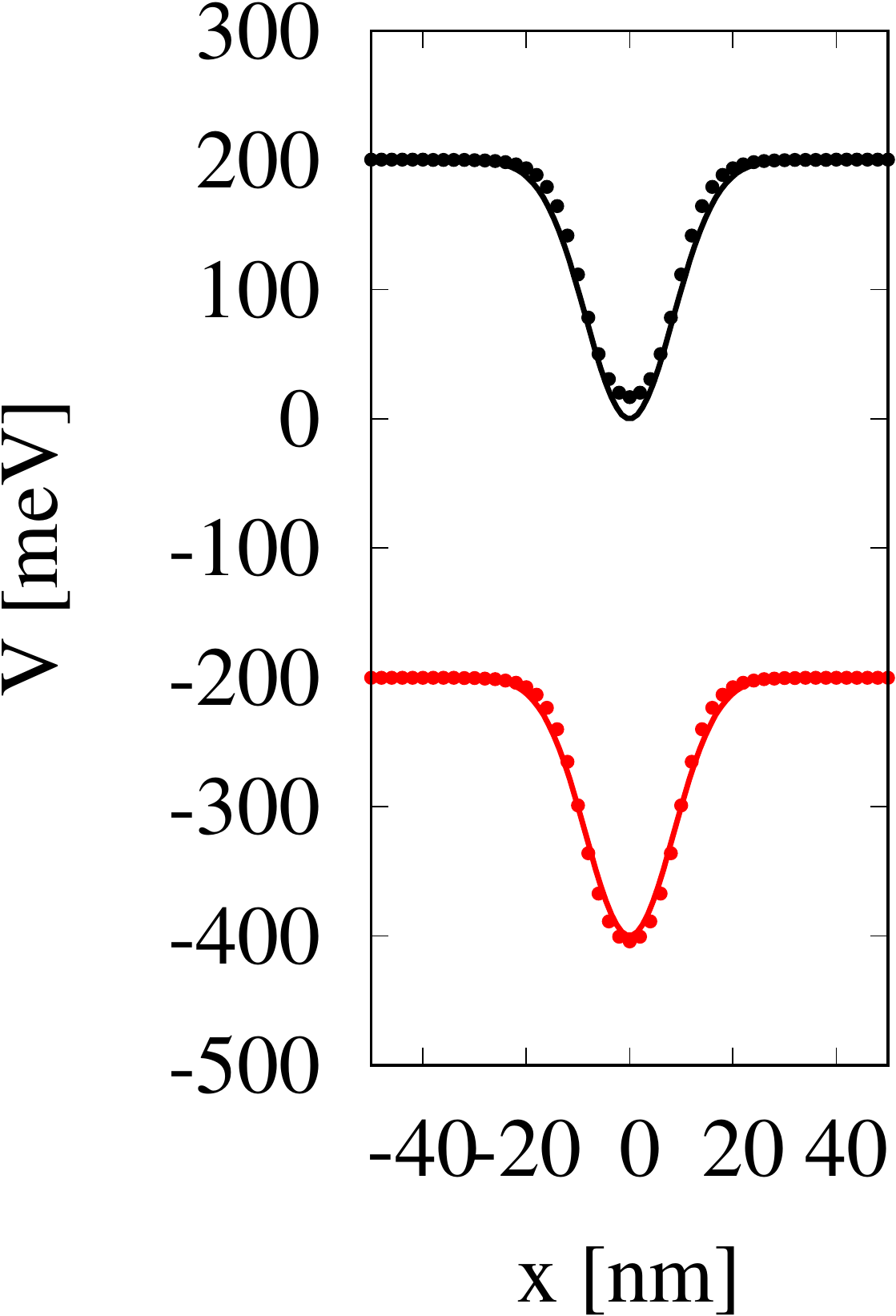} 
\end{tabular}
\caption{
(a) The geometry of the gated system that produces
a Gaussian confinement  within the silicene.
The silicence layer is embeded within a dielectric
sandwitched between two electrodes.
The two dotted lines indicate the A (lower line) and B (upper line) sublattices
of silicene that are shifted in the vertical direction by 
 by $\delta =0.046$ nm.  The midpoint between the sublattices is 
placed  a distance of  $H_S=0.5$ nm from the bottom electrode and $H=10$ nm from the top electrode.
 The top electrode has a circular protrusion of diameter $2R_p=20$ nm that is $H_p=0.5$ nm deep. 
(b) The symbols indicate the electrostatic  potential obtained  on the silicene sublattices for the bottom gate voltage $V_b=-3.8$ V and the top gate voltage $V_t=79.5$ V,
and the lines indicate the fit by Gaussian potentials 
$V_A=w-we^{-r^2/R_p^2}$, $V_B=-w-w e^{-r^2/R_p^2}$
for $w=200$ meV.
The black (red) color corresponds to the A (B) sublattice.
} \label{bb}
\end{figure}

The last term of Hamiltonian (\ref{hb0}) introduces the electrostatic potential
that should open the energy gap 
and  define the confinement potential.
In the calculations for a single quantum dot we model the  potential as due to the setup that is depicted in Fig. \ref{bb}(a). The silicene layer is sandwiched within a dielectric that separates two gates.
 The top gate contains a circular protrusion of radius $R_p$. 
Solution of the Laplace equation for the system of Fig. \ref{bb}(a) 
for $V_b=-3.8$ V and $V_t=79.5$ V
on the sublattices is presented in Fig. \ref{bb}(b).
The vertical electric field introduces potential difference between
the ions at the bottom (A, black line in Fig. \ref{bb}(b)) and
top (B, red line in Fig. \ref{bb}(b)) sublattice, and the difference opens
the energy gap \cite{ni,Drummond12,kiku,szafranre}.
The potential applied to the top gate is attractive for electrons
and its protrusion that is closer to silicene layer induces the dip of the confinement potential for the conduction-band electrons on both sublattices $A$ 
and $B$ [Fig. \ref{bb}(b)]. For the geometry assumed in Fig. \ref{bb}(b)
the potential has a form of a Gaussian 
 of nearly equal depth / width on both sublattices.
Based on this finding for $V_k$ in Eq. (\ref{hb0}) we set
\begin{equation}
V_k=V({\bf r_k})=\left\{\begin{array}{ll} V^A({\bf r}_k)= w-we^{-r_k^2/R_p^2} & \text{for $k$ in A} \\  V^B({\bf r}_k)= -w-we^{-r_k^2/R_p^2} & \text{for $k$ in B} 
 \end{array} \right., \label{numer}
\end{equation} 
with $R_p=10$ nm,   equal to the radius of the protrusion. 

\begin{figure}
\begin{tabular}{ll}
\begin{tabular}{l}
(a)  \includegraphics[width=0.4\columnwidth]{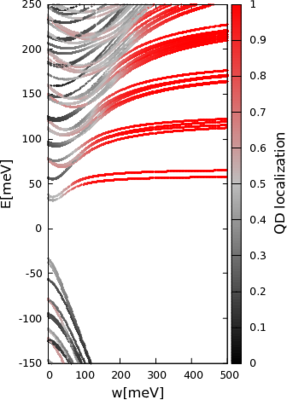}
\end{tabular} &
\begin{tabular}{l}
(b)  \includegraphics[width=0.55\columnwidth]{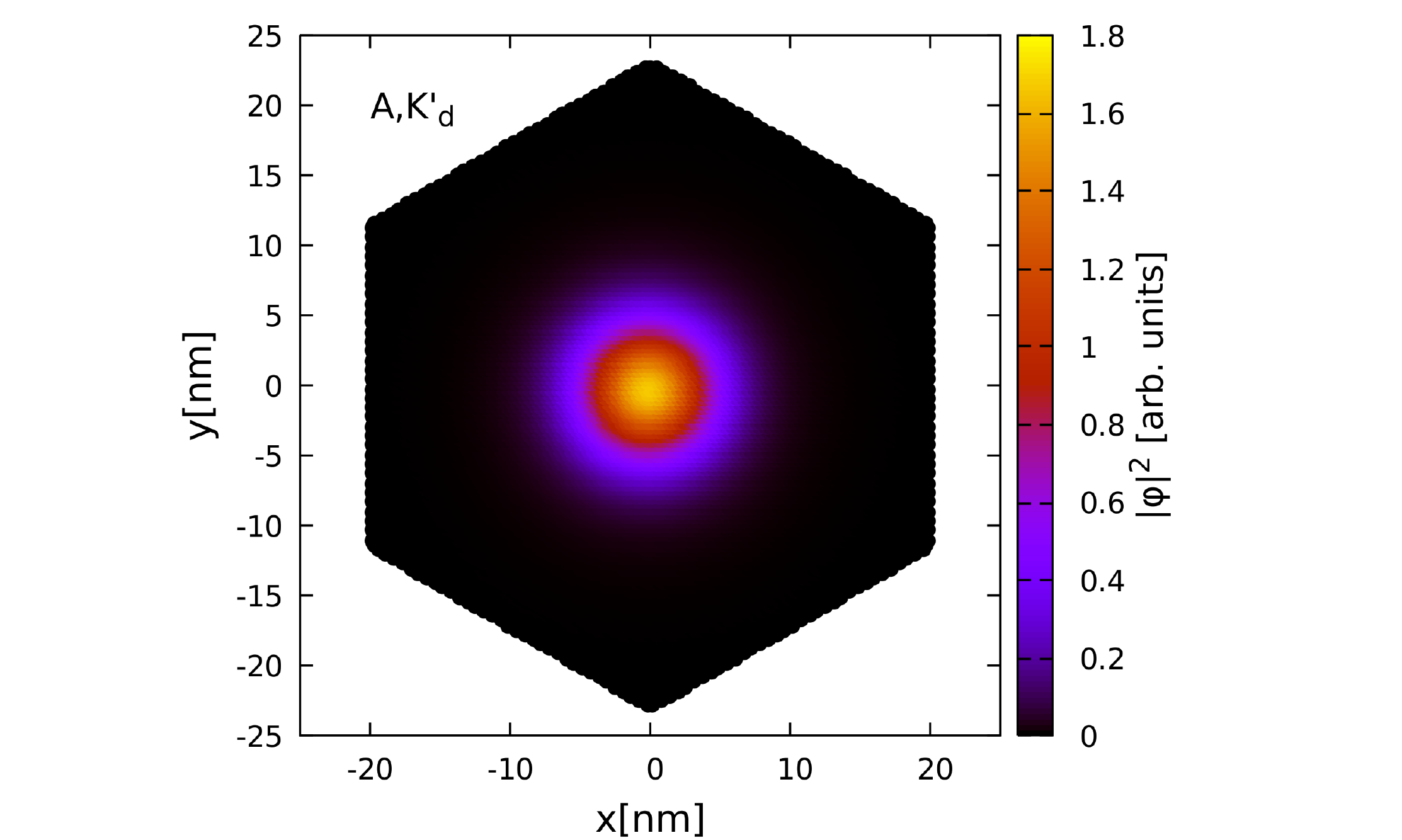}\\
(c) \includegraphics[width=0.55\columnwidth]{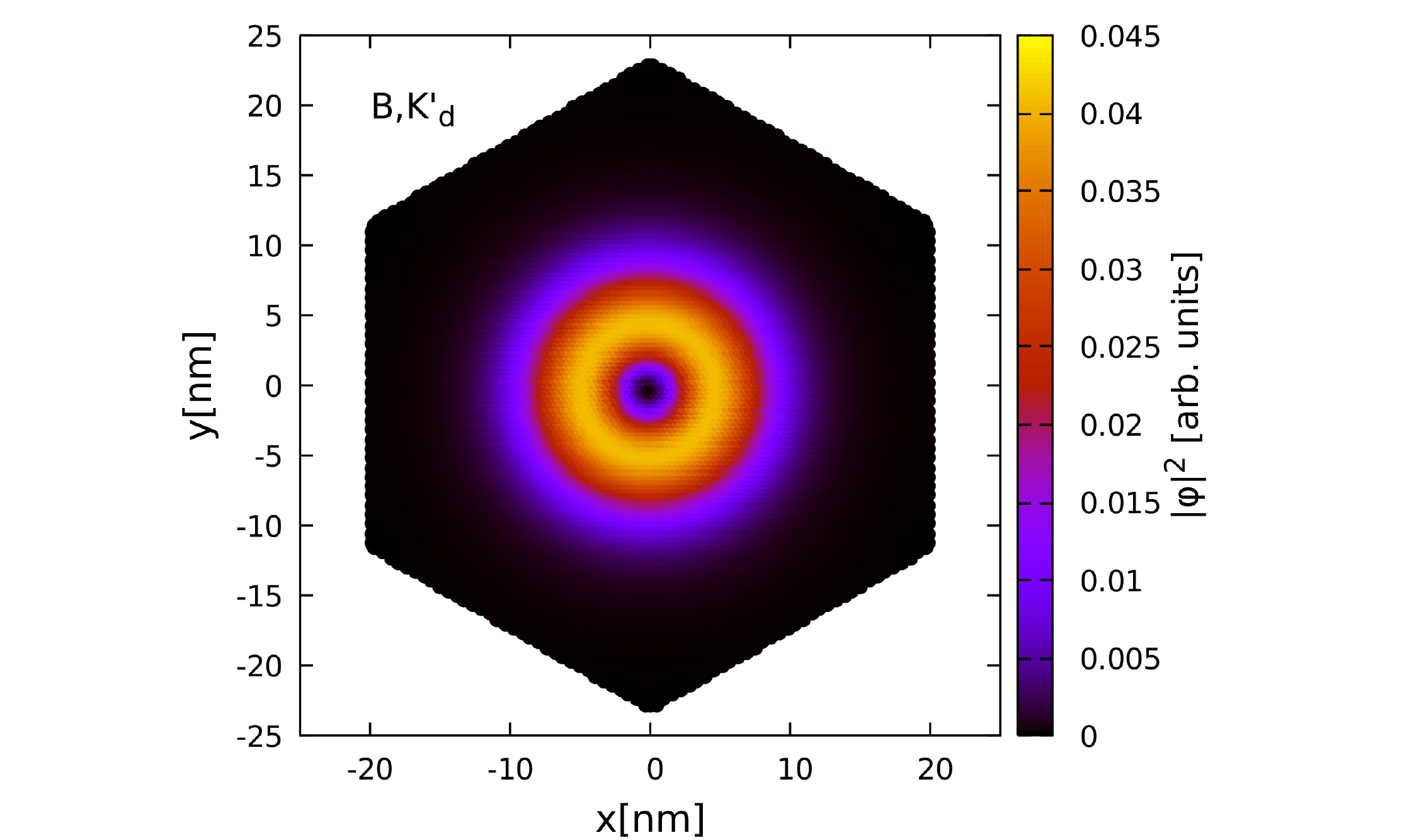}
\end{tabular}
\end{tabular}
\caption{(a) Energy levels as a function of the potential depth $w$. The colorscale indicates the localization of the energy levels in the quantum dot calculated as the integral of the 
charge density within the radius of 1.1$R_p$ from the dot center. The results are calculated in the absence of the external magnetic field. 
The conduction (valence) band states are found for $E>0$ ($E<0$).
(b-c) The $K'$ spin-down ground-state electron density 
on the A (b) and B (c) sublattices for $w=200$ meV. 
In the continuum approximation the angular momentum quantum number is 0 on the A sublattice and -1 on the B sublattice.
} \label{wiazanie}
\end{figure}

For the single quantum dot we consider a regular hexagonal flake 
with an armchair edge and  side length of 22 nm [Fig. \ref{wiazanie}(b,c)].
The positions of the ions of the A sublattice 
${\bf r}_{\bf k}^A=k_1 {\bf a}_1+k_2 {\bf a}_2$  are generated 
with the crystal lattice vectors
${\bf a}_1=a \left(\frac{1}{2},\frac{\sqrt{3}}{2},0\right)$
and ${\bf a}_2=a \left(1,0,0\right)$, where $a=3.89$ \AA\; is the silicene lattice constant, and $k_1$, $k_2$ are integers. 
The B sublattice ions are generated by ${\bf r}_{\bf k}^B={\bf r}_{\bf k}^A+(0,d,\delta)$,
with the in-plane nearest neighbor distance
 $d=2.25$ \AA\; and the vertical shift of the sublattices $\delta=0.46$ \AA.

\begin{figure*}
\begin{tabular}{lll}
(a) \includegraphics[width=0.42\columnwidth]{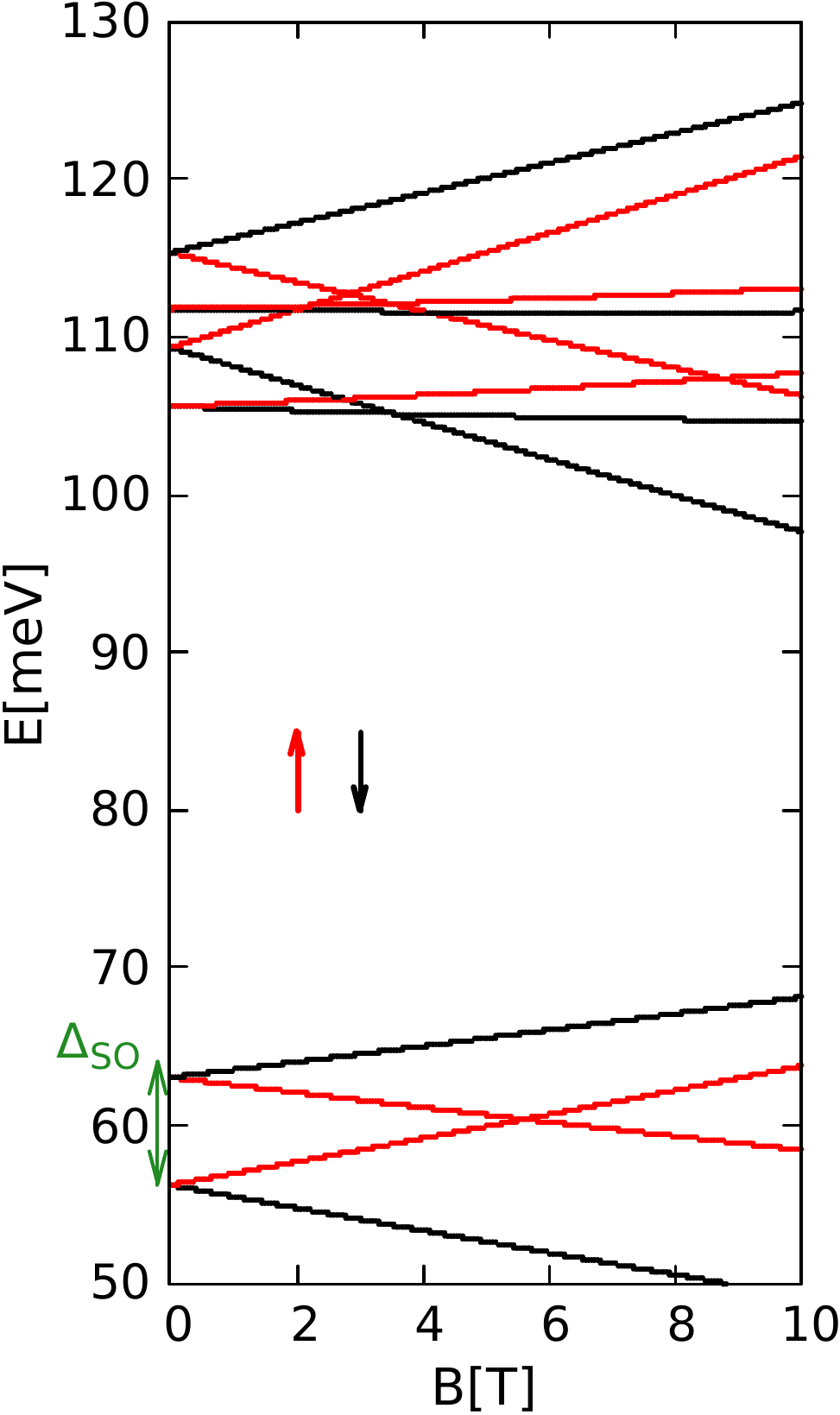} 
&
(b) \includegraphics[width=0.45\columnwidth]{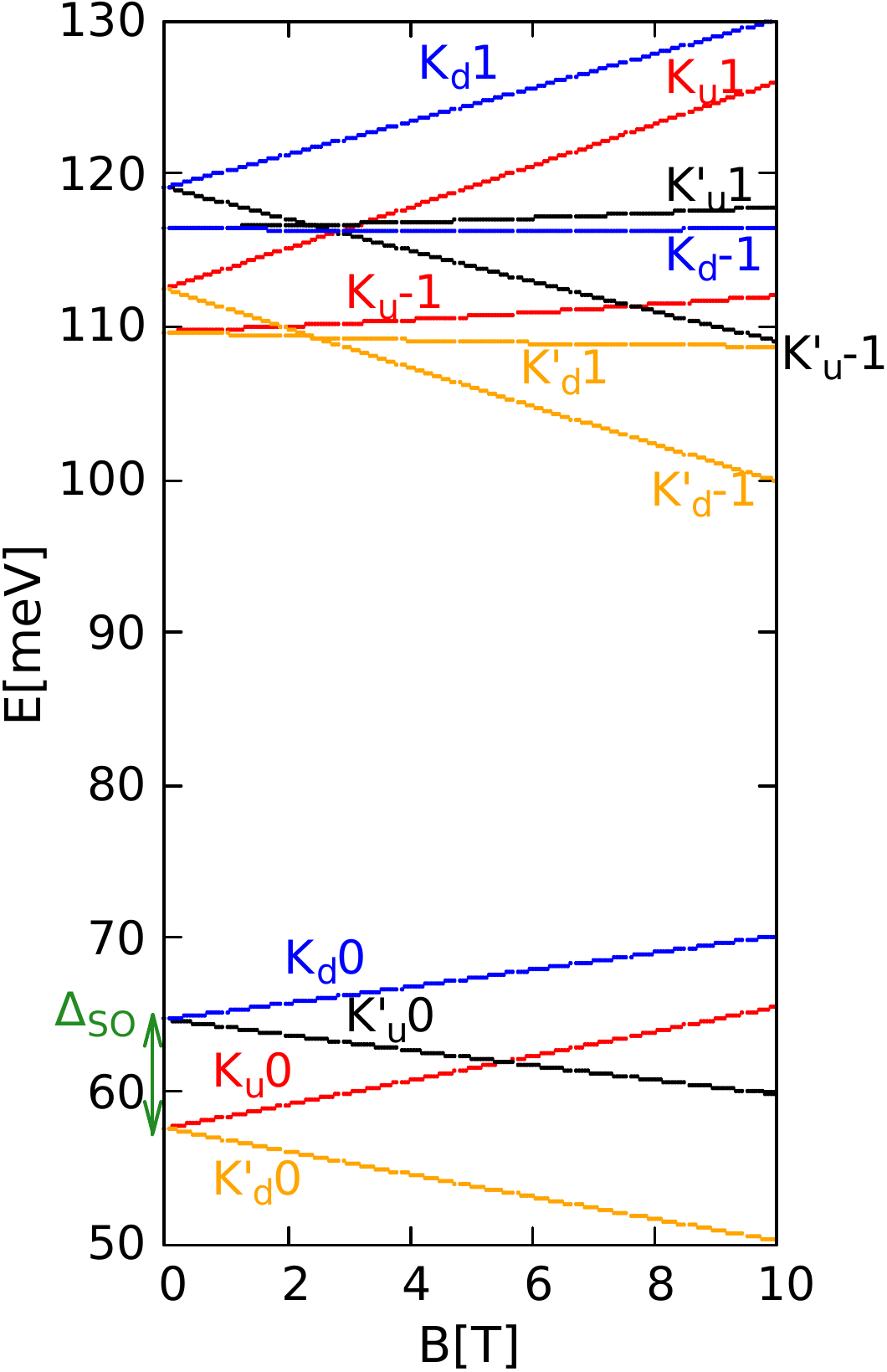} 
&
(c) \includegraphics[width=0.45\columnwidth]{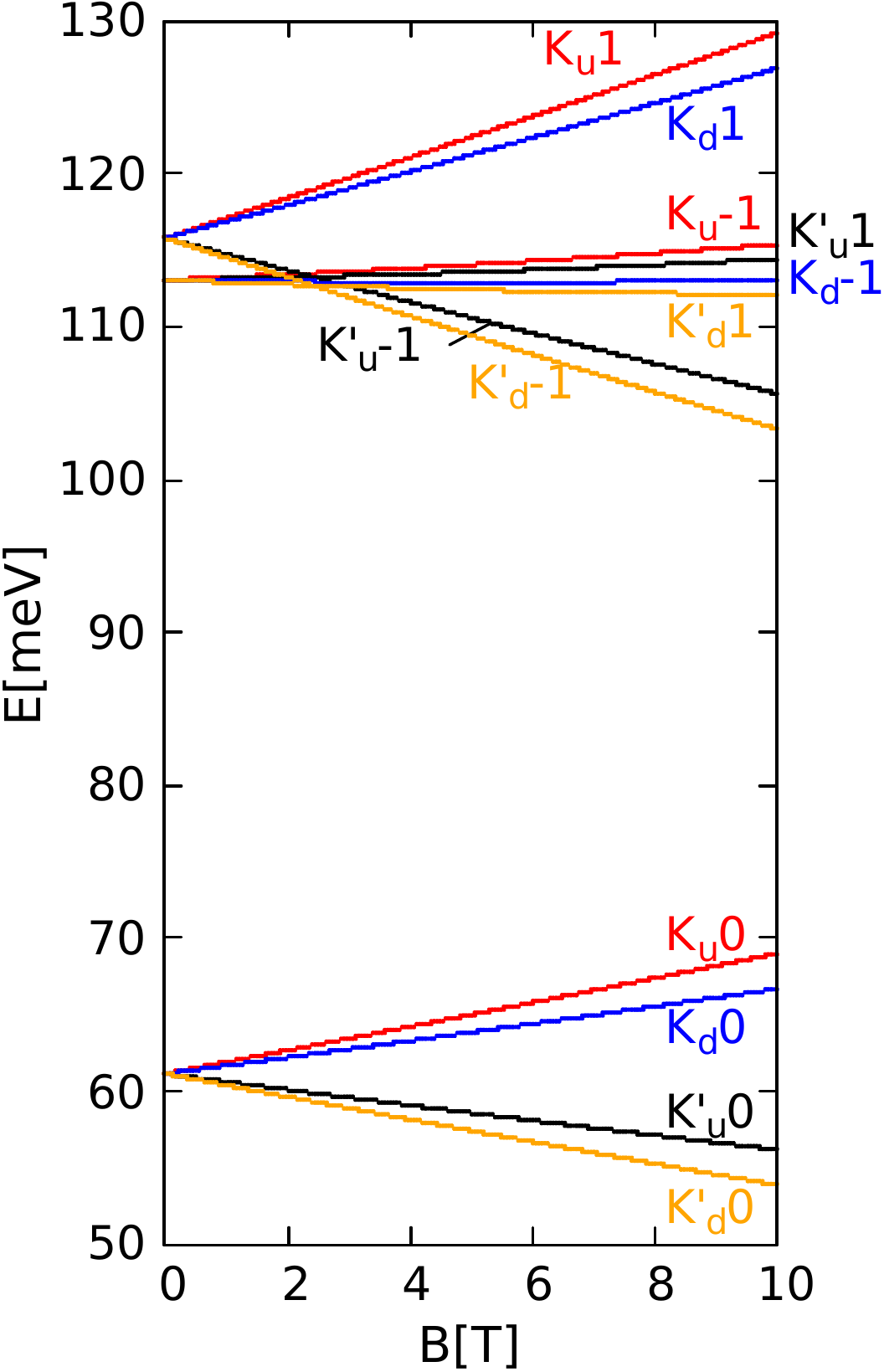} 
\end{tabular}
\caption{(a) The energy spectrum calculated by the atomistic tight-binding method. The black and red lines correspond to
the spin-down and spin-up states in the $s_z$ basis. 
(b,c) The energy spectrum of the continuum Hamiltonian. The energy levels 
are labeled by the valley index $K$ or $K'$. The spin in the $s_z$
basis is marked by the $u$ or $d$ subscript for the spin-up
and spin-down orientation, respectively. The integer after the valley index stands
for the angular momentum quantum number for the wave function
on the A sublattice. In (c) the spin-orbit interaction is neglected. In (b) the intrinsic spin-orbit interaction is present. In (a) both the intrinsic spin-orbit interaction 
and the Rashba coupling are present. The influence on
the latter on the energy spectrum is negligible.
The results were obtained for $w=333$ meV, respectively.
%$\Delta_{SO}$ marks the ground-state spin-orbit splitting.
} \label{widma1eb}
\end{figure*}
\subsection{Electron confinement}
The energy levels calculated with the tight-binding
Hamiltonian are displayed as a function of the
potential depth $w$ in Fig. \ref{wiazanie}(a). 
The colorscale in Fig. \ref{wiazanie}(a) shows the 
localization of the eigenstates within the range of 1.1 $R_p$ from the center of the electrostatic potential. 
The localized states are found only in the conduction-band side ($E>0$)  of the spectrum. 
The states of the valence band are localized outside of the dot.
The energy gap rapidly increases with $w$. Namely, for $w=100$ meV, 200 meV
and 300 meV  the gap is 170 meV, 290 meV and 380 meV, respectively.

The wave function of the spin-down ground state 
is given for $w=200$ meV in Fig. \ref{wiazanie}(b,c). 
The electron density of the localized conduction band
states is nearly entirely localized at the sublattice $A$.

\subsection{Magnetic field}

The external magnetic field is introduced by the  Peierls phase
via modification of the hopping terms $c^\dagger_{k\alpha}c_{l\beta}\rightarrow
c^\dagger_{k\alpha}c_{l\beta}  e^{i\frac{e}{\hbar}\int_{\vec{r_k}}^{\vec{r_l}}\vec {A}\cdot \vec {dl}}$ with the vector potential $\vec{A}$. For the perpendicular magnetic field  $\vec{B}=(0,0,B)$ we use the  symmetric gauge $\vec{A}=(-Bx/2,By/2,0)$.
With the Peierls phase included the energy operator changes to $H_0\rightarrow H_b$, 
and the Hamiltonian is completed by the spin Zeeman term
\begin{equation}
H_B=H_b+\frac{1}{2}g\mu_B B\sum_{k,\alpha}  \sigma^z_{\alpha,\alpha} c_{k\alpha}^\dagger c_{k\alpha},
\end{equation}
where 
 $\mu_B$ is the Bohr magneton $g=2$ and is the Land\'e factor.

\subsection{Continuum approximation}

When the electron system gets localized inside the quantum dot, 
the intervalley mixing by the edge \cite{zarenia} becomes negligible. 
Then, the valley index becomes a good quantum number.
In order to identify the valley we perform calculations
using the continuum approximation \cite{Ezawa12a} of the tight-binding Hamiltonian (\ref{hb0}) that keeps track of the diagonal intrinsic spin-orbit interaction.
For the identification of the quantum numbers we neglect
the Rashba coupling that, although  crucial for spin manipulation,
 produces only a slight modification to the energy levels and majority-spin eigenfunctions. 
For the wave function with two components, each corresponding
to a single sublatice  $\psi=\left(\begin{matrix} \psi_A \\ \psi_B\end{matrix}\right)$,
the continuum Hamiltonian reads \cite{Ezawa12a}
\begin{equation} H_\eta = \hbar V_F \left(k_x \tau_x -\eta k_y \tau_y \right)+V({\bf r})\tau_z+\frac{g\mu_B B}{2}\sigma_z -\eta \tau_z \sigma_z 3\sqrt{3} t_2,\end{equation}
where the valley index is $\eta=1$ for the $K$ valley and $\eta=-1$  for the $K'$ valley, 
and $\tau_x$, $\tau_y$ and $\tau_z$ are the Pauli
matrices in the sublattice space, ${\bf k}=-i\nabla+\frac{e}{\hbar} \vec{A}$, and $V_F=\frac{{3dt}}{2\hbar}$ is the Fermi velocity.
For the circular potential $V(r)$,  the $H_\eta$ Hamiltonian eigenstates $\Psi_\eta$
are also eigenstates of 
the valley-orbital angular momentum operator of form $J_z=L_z \left(\begin{matrix}1&0 \\ 0 & 1\end{matrix}\right) +\eta \frac{\hbar}{2} \tau_z$, where $L_z=-i\hbar \frac{\partial }{\partial \phi}$ stands for the operator of the orbital angular momentum.
\begin{equation} \Psi_\eta=\left(\begin{matrix}  f_A (r) \exp(il\phi) \\ f_B (r) \exp(i(l+\eta)\phi)\end{matrix} \right) \label{psi} \end{equation}
where $l$ is the orbital quantum number,
and 
the eigenequation for the radial functions reads
\begin{eqnarray}
&& (V^A(r)-\eta 3\sqrt{3}t_2  \sigma_z +\frac{g\mu_B B \sigma_z}{2})f_A\nonumber \\ &&+V_F\left[-\eta\frac{i\hbar}{r}(l+\eta)f_B-i\hbar f'_B-\eta \frac{iBr}{2}f_B \right]=E f_A, \label{e1}\\
&& (V^B(r)+\eta 3\sqrt{3}t_2  \sigma_z +\frac{g\mu_B B \sigma_z}{2})f_B\nonumber \\ && +V_F\left[ \eta\frac{i\hbar}{r}l f_A-i\hbar f'_A+\eta \frac{iBr}{2}f_A \right]=E f_B. \label{e2}
\end{eqnarray}
The system of eigenequations (\ref{e1},\ref{e2}) is solved with a finite difference method.

\subsection{Electron pair}

%Generally in graphene flakes a good quantum number is 
%the excess charge only and not
%the number of carriers in the conduction and valence bands.
In graphene flakes, the Coulomb interaction of the excess electrons
can be strong enough to 
 generate an extra electron and hole pair \cite{egger}.
The present electrostatic quantum dot attracts the electrons but repulses the holes
and the confinement potential 
does not support  localized hole states of the valence band [Fig. \ref{wiazanie}(a)].
The cost of generation of the electron-hole pair 
is of the order of the energy gap, e.g. 290 meV for $w=200$ meV.
Even for the small size of the present dot  --  the electron-electron
interaction energy is eight times smaller -- about $36$ meV (for $w=200$ meV). 
For a pair of confined conduction band electrons 
the generation of extra electron-hole pair will not decrease
the electron-electron repulsion, since the extra electron is added
to the dot, and the hole needs to stay outside. 
For that reasons below we fix the number of conduction band electrons \cite{c2} to two
and consider the Hamiltonian  
\begin{equation}
  {H}_{2e}=\sum_{i}{d}^{\dagger}_{i}{d}_{i}E_i +\frac{1}{2}\sum_{ijkl}{d}^{\dagger}_{i}{d}^{\dagger}_{j}{d}_{k}{d}_{l}V_{ijkl},
  \label{Htwo}
\end{equation}
where ${d}^{\dagger}_{i}$ is the electron creation operator for the single-electron energy level $E_i$  and the Coulomb matrix elements read
\begin{equation}
V_{ijkl}=\kappa\langle{\psi_{i}(\mathbf{r_1})\psi_{j}(\mathbf{r_2})\frac{1}{|\mathbf{r_{12}}|}}{\psi_{k}(\mathbf{r_1})\psi_{l}(\mathbf{r_2})}\rangle, 
\label{coulF}
\end{equation}
where $\kappa=e^2/(4\pi\epsilon\epsilon_0)$ and we use the Al$_2$O$_3$ dielectric constant  $\epsilon_0=9.1$.
We integrate the Coulomb elements for the single-electron wave functions $\psi$ spanned by the
Si atomic orbitals $3p_z$,
\begin{equation}
\psi_i({\bf r}_i)=\sum_{k,\sigma_k} C^i_{k,\sigma_k} p_z^k({\bf r}_1),
\end{equation}
where the summation over spin accounts for the effects of the non-spin-diagonal Rashba interaction. 
Although the spins are nearly polarized perpendicular to the silicene plane
by the spin-diagonal intrinsic-spin-orbit interaction, the trace contribution of the minority
spins  allows for the spin transitions. 
The Coulomb matrix element reads 
\begin{eqnarray}
V_{ijkl}&=&\kappa\langle\psi_{i}({\bf r_{1}})\psi_{j}({\bf r_{2}})|\frac{1}{|{r_{12}}|}|\psi_{k}({\bf r_{1}})\psi_{l}({\bf r_{2}})\rangle\nonumber \\
&=&\kappa\sum_{\substack{a,\sigma_{a};b,\sigma_{b};\\ c,\sigma_{c};d,\sigma_{d} }}C_{a,\sigma_{a}}^{i*}C_{b,\sigma_{b}}^{j*}C_{c,\sigma_{c}}^{k}C_{d,\sigma_{d}}^{l}\delta_{\sigma_{a};\sigma_{d}}\delta_{\sigma_{b};\sigma_{c}}\times \nonumber\\&&\langle p_{z}^{a}({\bf r}_{1})p_{z}^{b}({\bf r_{2}})|\frac{1}{|{r_{12}}|}|p_{z}^{c}({\bf r_{1}})p_{z}^{d}({\bf r_{2}})\rangle.
\end{eqnarray}
For the Coulomb integral we apply the two-center approximation \cite{c2}
$ \langle p_{z}^{a}({\bf r}_{1})p_{z}^{b}({\bf r_{2}})|\frac{1}{|{r_{12}}|}|p_{z}^{c}({\bf r_{1}})p_{z}^{d}({\bf r_{2}})\rangle=\frac{1}{r_{ab}} \delta_{ac}\delta_{bd}$ for $a\neq b$. 
The  on-site integral ($a=b$) is calculated with the $3p_z$  Si atomic orbitals,  $p_z({\bf r})=N z \left(1-\frac{Z r}{6}\right)\exp(-Zr/3)$, where $N$ is the normalization constant and $Z$ is the effective screened Si nucleus charge as seen by $3p_z$ electrons. The single-center integral can then be 
calculated analytically and is equal to $I_{a=b}=\frac{3577}{46080}Z$. The Slater screening rules for $3p$ electrons produce $Z=4.15$, then $I_1=8.76$ eV.

The Hamiltonian (\ref{Htwo}) is diagonalized with the configuration interaction approach in the basis of up to $\sim 3000$ two-electron Slater determinants
constructed from the lowest-energy  eigenfunctions
of the single-electron Hamiltonian (3). 

\subsection{Driven transitions}

We study the charge, spin and valley dynamics for 
the system subject to an in-plane  AC electric field.
The time-dependent Hamiltonian for the field oriented along the $y$ axis reads 
\begin{equation} H_t=H_B+H'=H_B+eF_{ac}\sum_{k,\alpha}y_k\sin(2\pi\nu t) c_{k\alpha}^\dagger c_{k\alpha}.
\end{equation}
The amplitude and the frequency of the AC electric field are denoted by $F_{ac}$ and $\nu$, respectively. 
For integration of the time dependent  Schr\"odinger equation $i\hbar \frac{\partial \Psi}{\partial t}=H_t \Psi$, 
we use the eigenstates of the stationary Hamiltonian ($H_B\psi_n=E_n\psi_n$), 
\begin{equation} \psi=\sum_n A_n(t) \exp(-\frac{iE_n t}{\hbar})\psi_n, \label{basis} \end{equation}
which gives the following system of equations
\begin{equation}
i\hbar \frac{dA_k(t)}{dt}=\sum_n A_n(t) eF_{ac} \sin(2\pi\nu t) y_{kn} e^{-i\frac{E_n-E_k}{\hbar}t}, \label{time}
\end{equation}
with the dipole matrix elements $y_{kn}=\langle\Psi_k | y|\Psi_n\rangle$. 

For two electrons the system of equations for description of the time-dependence  is formally identical,
only with $E_n$ standing for the two-electron eigenergies and the matrix elements $y_{kn}$ calculated
for the two-electron wave functions. In the calculations we set the stationary ground-state
in the initial condition and study the system dynamics for the AC pulse duration of $3.74$ ns. 
%We consider $F_{AC}$ 2$ kV/cm and $F_{AC}=0.2$ kV/cm 

\section{A single quantum dot}
\subsection{Single-electron states}

The splitting of the energy levels in Fig. \ref{wiazanie}(a)
at large $w$
is due to the intrinsic spin-orbit coupling. Each
of the energy levels is two-fold degenerate. The structure of the low-energy spectrum
for the conduction band states localized in the dot is
displayed in Fig. \ref{widma1eb}.
Figure \ref{widma1eb}(a) shows the results of the atomistic
tight-binding calculations with the color of the lines
showing the spin-up and spin-down states. 
In Fig. \ref{widma1eb}(b) we plotted, for comparison,
the results of the continuum approach. 
In Fig. \ref{widma1eb}(b) the levels are labelled 
by the valley index $K$ or $K'$, the spin $u$ or $d$
in the subscript for the spin-up and spin-down states,
and the integer shows the angular momentum quantum number
on the A sublattice. For comparison Fig. \ref{widma1eb}(c) 
indicates the results without the spin-orbit coupling,
when at $B=0$ each energy level is degenerate with 
respect to the valley and the spin. 
The intrinsic spin-orbit coupling introduces
the valley-spin interaction that splits the fourfold 
degenerate states to spin-valley doublets.
The states of opposite valleys produces electrical currents of opposite
orientations. The lower-energy (higher) doublets  correspond to 
parallel (antiparallel) orientation of the orbital and spin magnetic moments.

\begin{figure*}
\begin{tabular}{lll}
(a) \includegraphics[width=0.45\columnwidth]{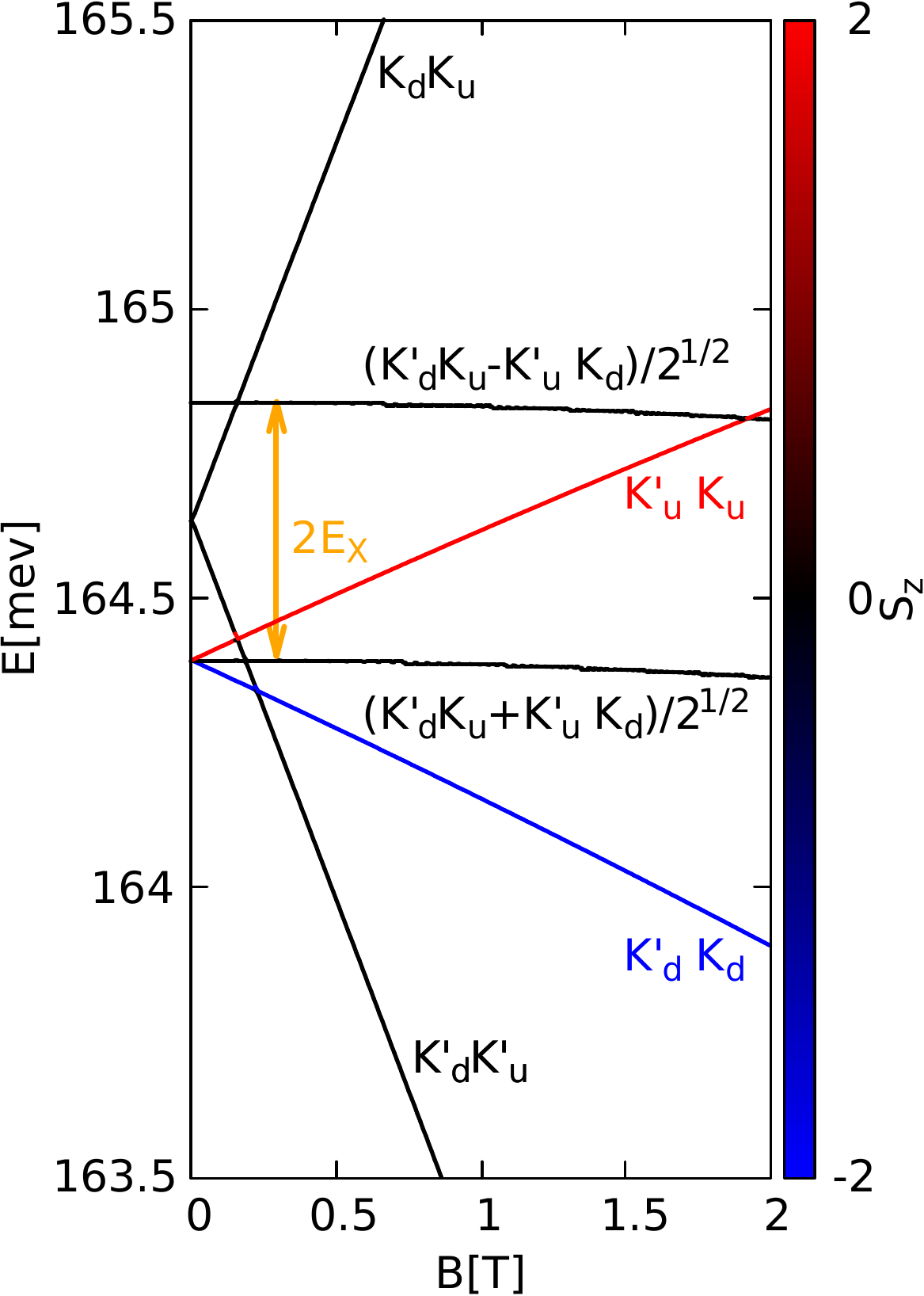} &
(b) \includegraphics[width=0.45\columnwidth]{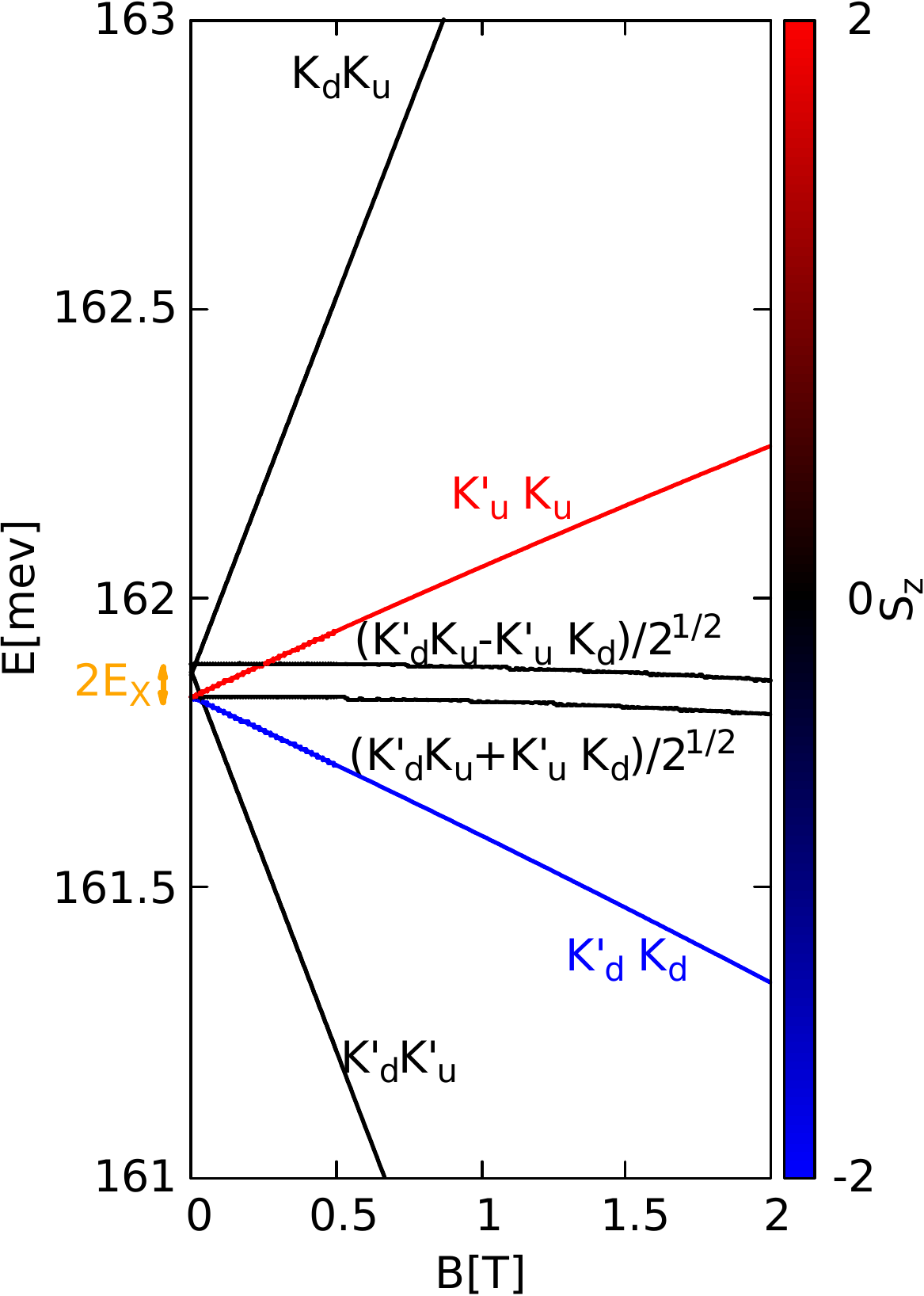} &
(c) \includegraphics[width=0.43\columnwidth]{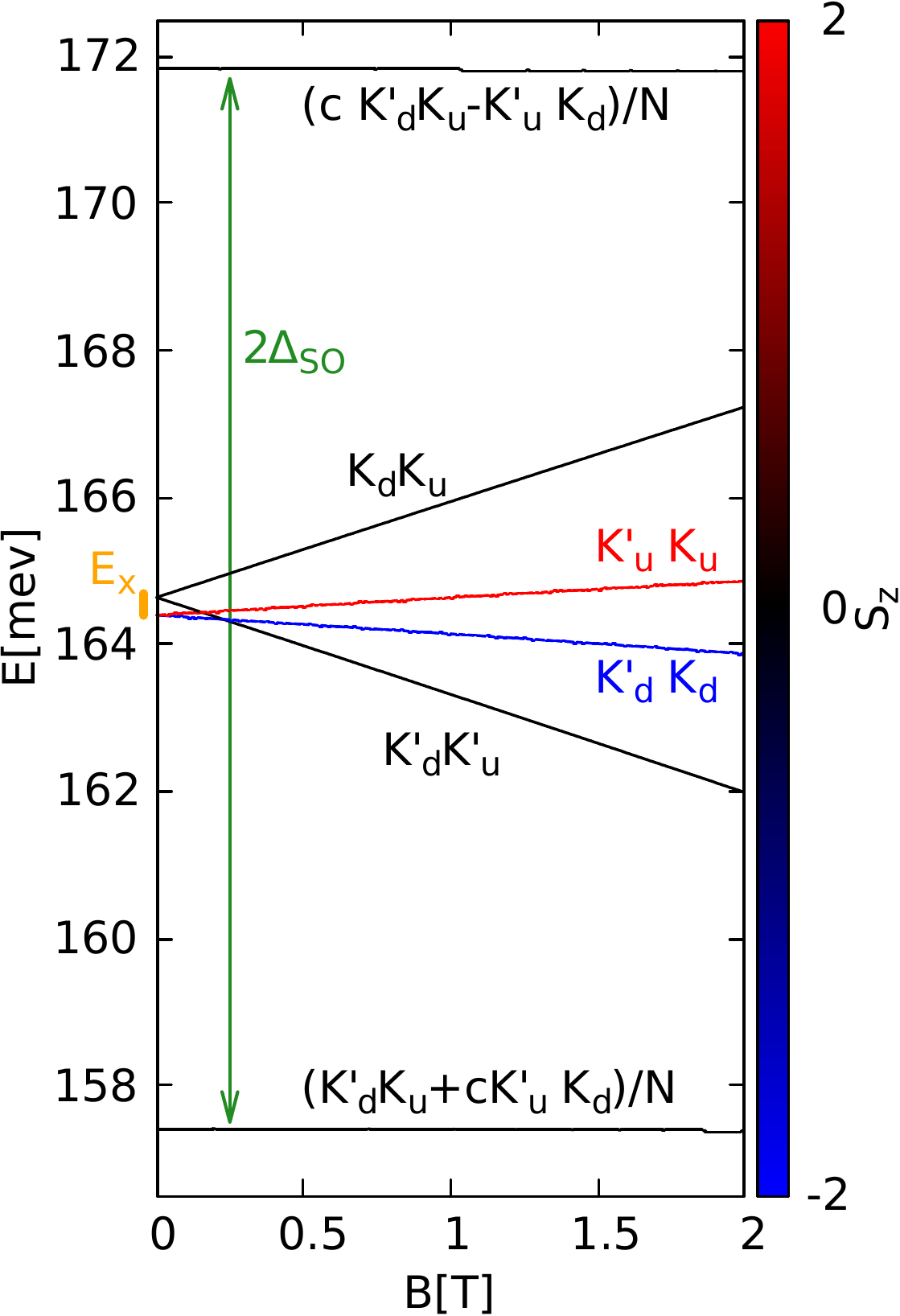} 

\end{tabular}
\caption{ The energy spectrum for a pair of electrons
in the quantum dot with $w=333$ meV. 
The results without (a,b) and with (c) the spin-orbit coupling. The value of $E_x$ stands for the intervalley exchange energy (see text). 
In (a) and (c) the calculations provide $E_x=0.22$ meV. 
In (b) the maximal value of the $1/r_{12}$ term in the integrand was set to $1/(3a)$, where $a$ is the lattice constant. The cut-off of the shortest-range part of the
Coulomb interaction reduces  $E_x$ in (b) to $0.029$ meV.
In the figure we indicate the occupation of the single-electron levels, where e.g. $K_dK_u$ stands for the two-electron Slater
determinant constructed from $K$ valley spin-down and spin-up energy levels etc.  In (c) the contribution
of the $K'_u K_d$ to the $K'_dK_u$ ground state is about 4\% in 
terms of the probability density. 
} \label{ep}
\end{figure*}

\begin{figure}
\begin{tabular}{ll}
(a) \includegraphics[width=0.44\columnwidth]{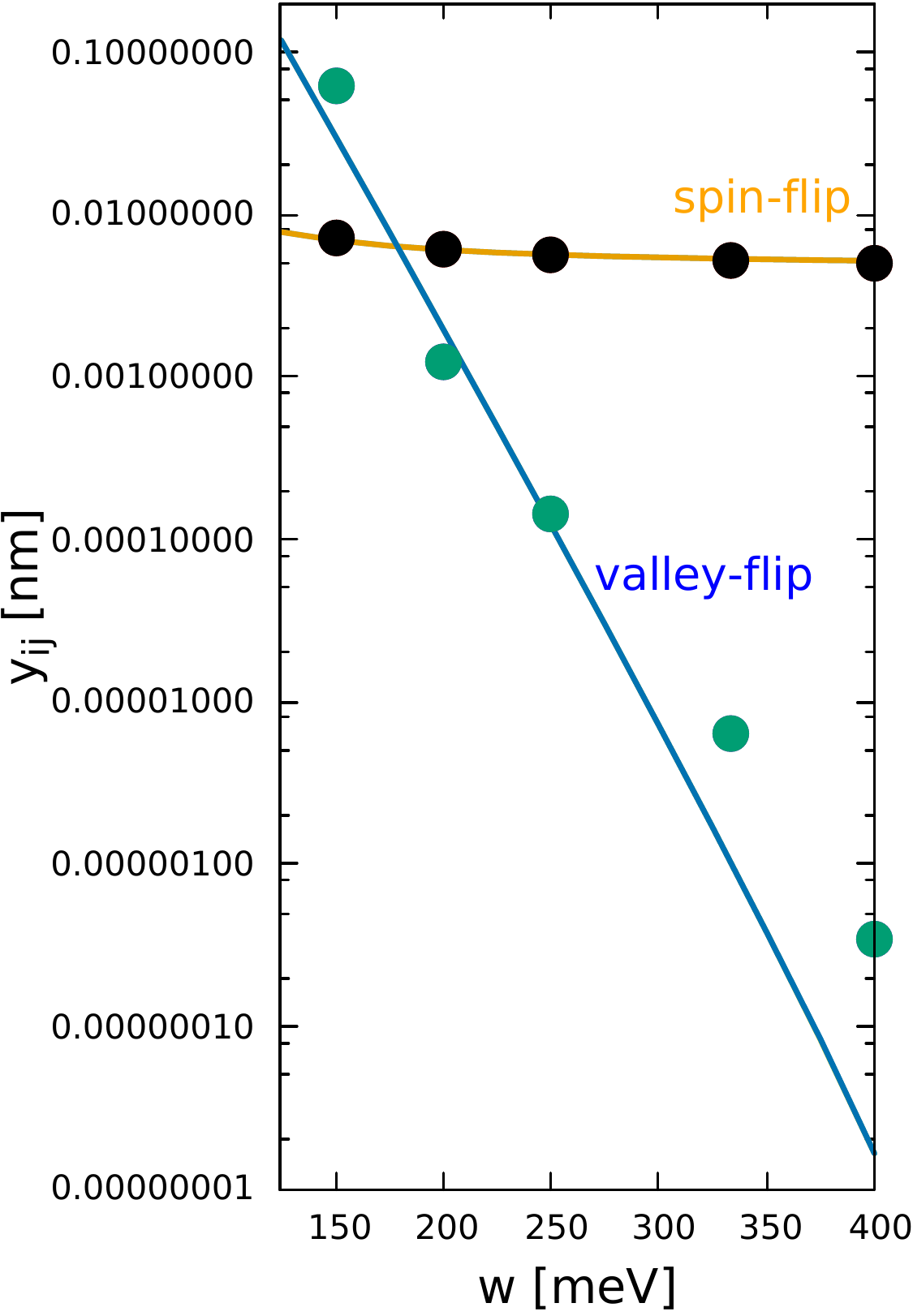} &
(b) \includegraphics[width=0.425\columnwidth]{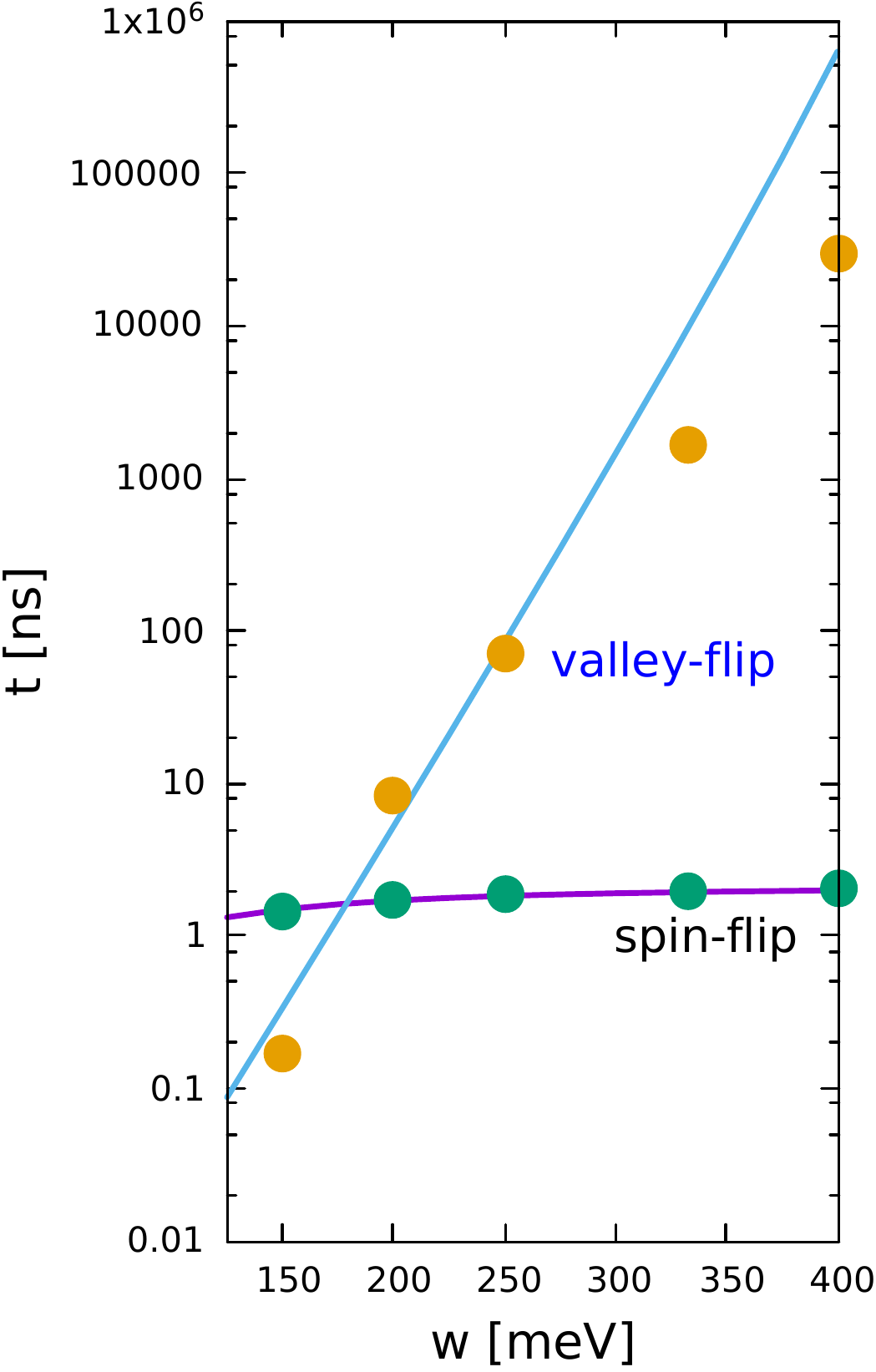}  
\end{tabular}
\caption{(a) Solid lines show the  transition matrix elements $y_{ij}=|\langle \phi_i | y | \phi_j \rangle |$ 
for a single excess electron confined in the quantum dot.
For the initial wave function $\phi_i$ we take the ground state $K'_d$ wave function at 1 T. The spin-flipping matrix element is calculated for 
the $\phi_j$ set as $K'_u$ The valley transition matrix element is obtained for $\phi_j$ set as $K_d$ energy level
-- the highest energy level for the quadruple of lowest-energy conduction band states in Fig. 3(b).
The points correspond to the results obtained for two confined electrons. The results are obtained for
the ground-state set as $\phi_i$ [see the spectrum Fig. 4(c)]. The valley (spin)  flipping transition obtained for $\phi_j$  identified with the first [black line in Fig. 4(c)]
(second [blue line in Fig. 4(c)] energy level wave function. 
The matrix elements for both spin and valley flips are 
to small to enter the plot. 
(b) Same as (a) only for the transition times obtained
for the AC electric field amplitude of $F_{AC}=2$kV/cm,
as $0.103/y_{if}$[nm $\times$ ns].
}
 \label{tramael}
\end{figure}

\begin{figure}
\begin{tabular}{l}
(a) \includegraphics[width=0.75\columnwidth]{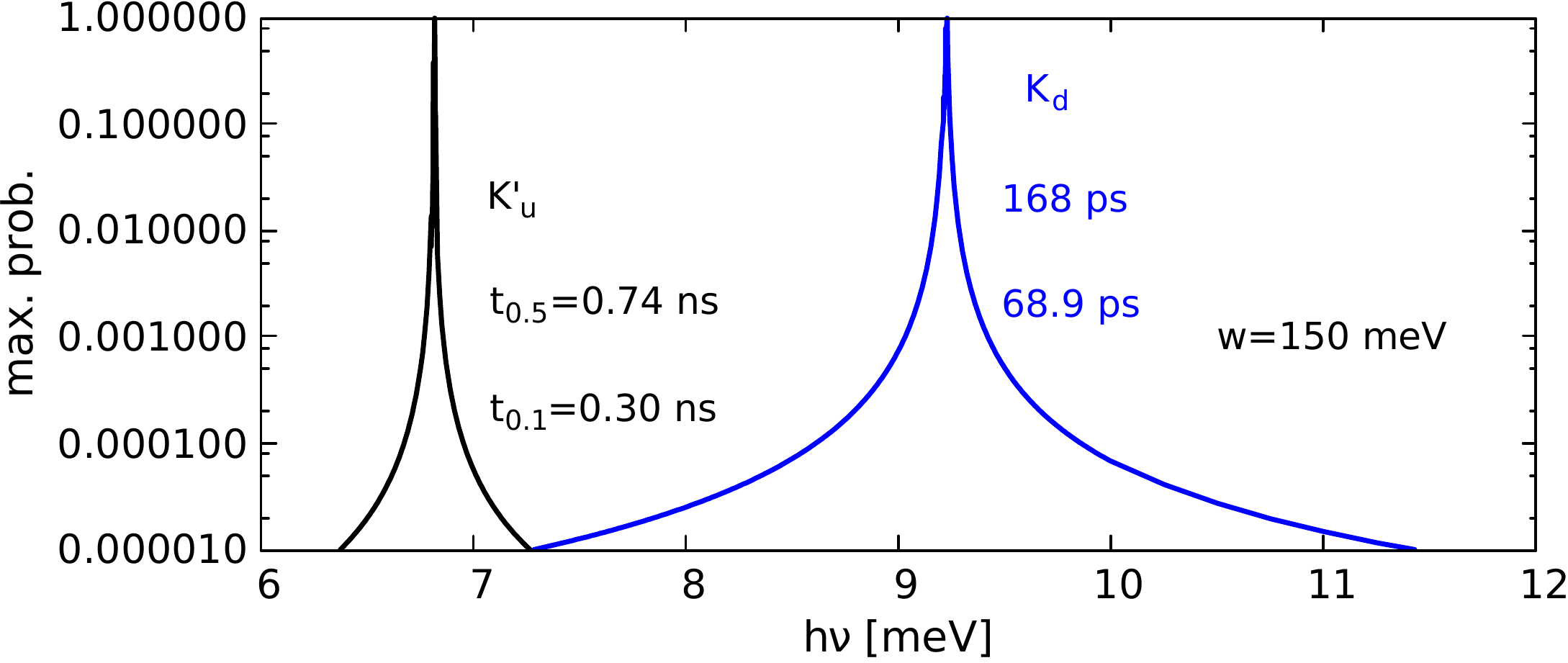} \\
(b) \includegraphics[width=0.75\columnwidth]{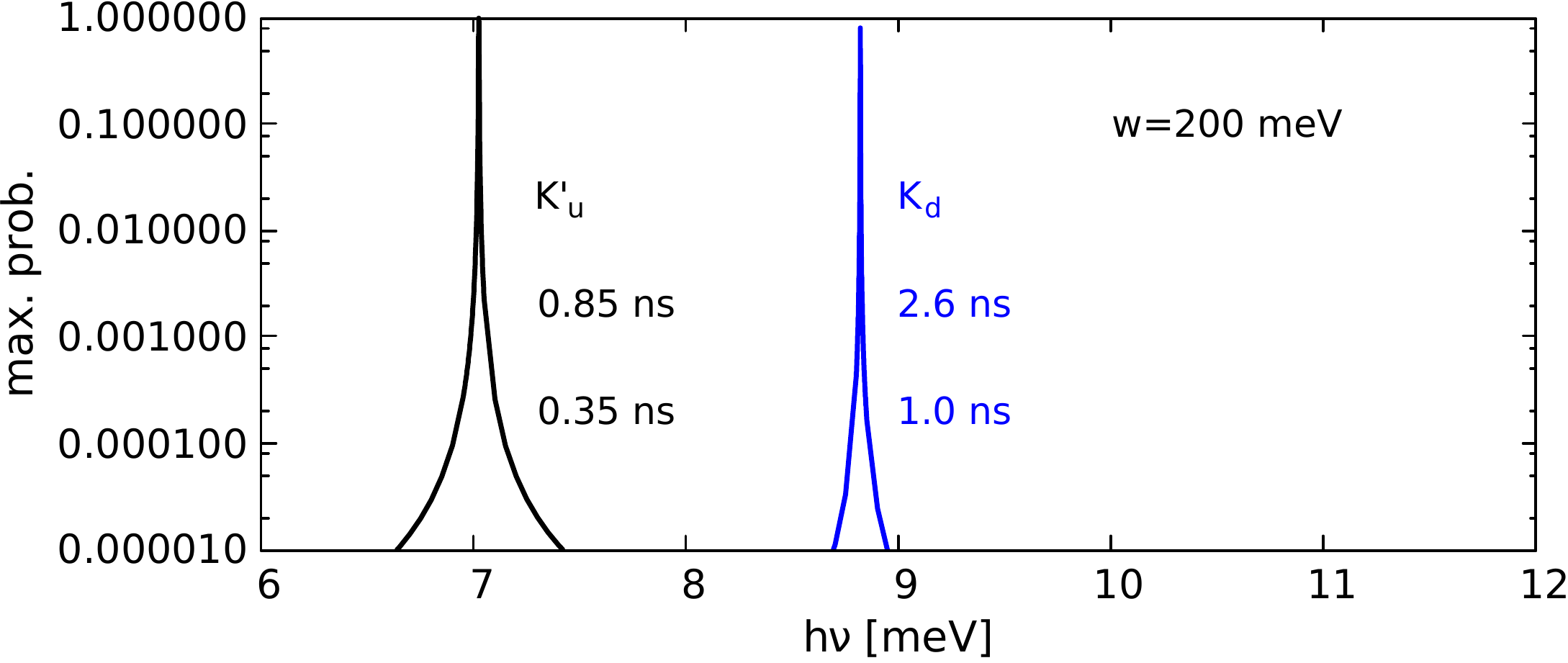}   \\
(c) \includegraphics[width=0.75\columnwidth]{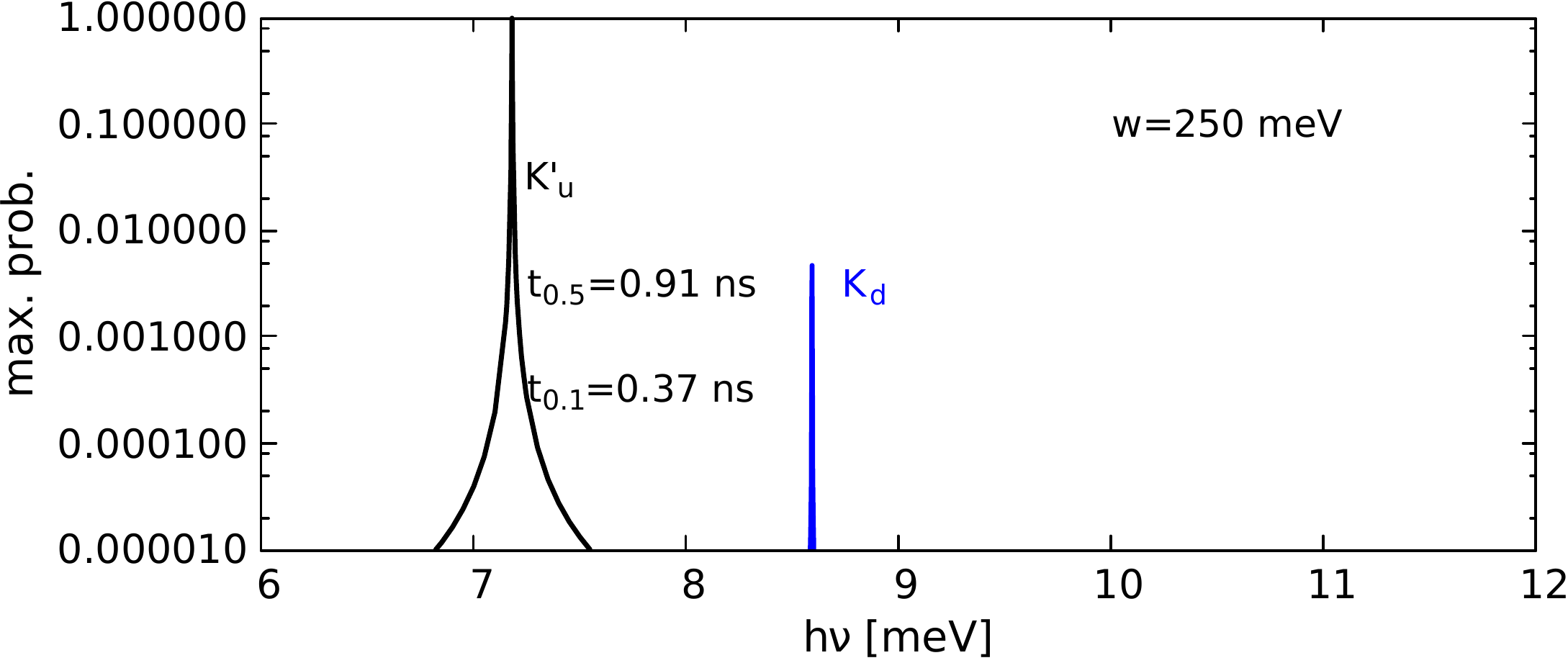}  \\
(d) \includegraphics[width=0.75\columnwidth]{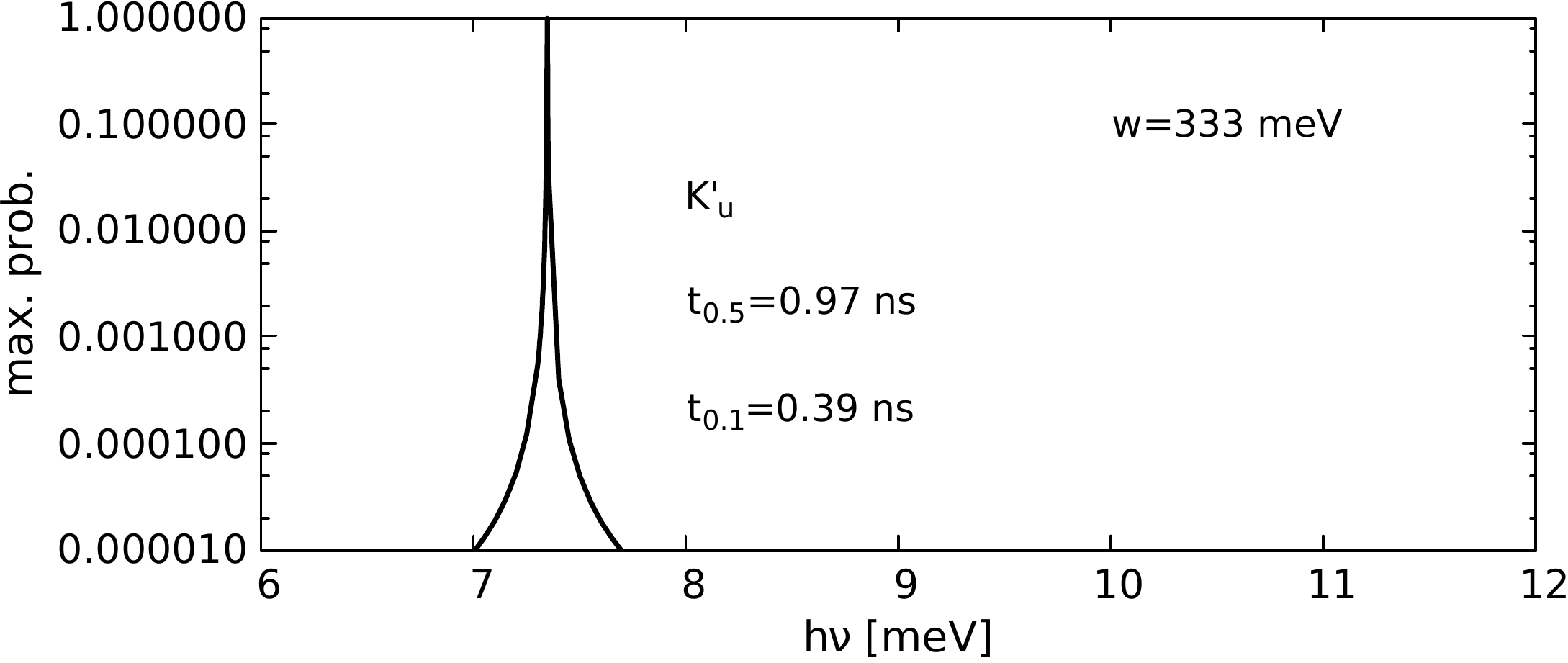}  \\
\end{tabular}
\caption{
Results for the time-dependent simulation lasting 3.74 ns
for the AC electric field given by $eF_{AC}y\sin(h\nu t)$. 
The single excess electron in the quantum dot is initially 
set in the $K'_d$ stationary ground state at the magnetic field of 1T. The plots indicate the maximal occupation of $K'_u$ and $K_d$ stationary excited energy levels obtained during the simulation. The numbers near the peaks indicate the duration of the $F_{AC}$ pulse upon which the probability
of finding the electron in the $K'_u$ (black lines) or $K_d$ (blue lines) energy levels 
exceeds 10\% ($t_{0.1}$) and 50\% ($t_{0.5}$). 
The spin and valley transition to the remaining energy level
of the quadruple $K_u$ -- is too small to fit in the plot.
Plots from (a) to (d) correspond to $w=150$ meV, $w=200$ meV,
$w=250$ meV and $w=333$ meV.
}
 \label{czas1}
\end{figure}

\subsection{The electron pair}

Near the ground-state each of the two electrons 
occupies  the states of the two lowest-energy single-electron doublets.
In consequence, the two-electron ground-state is nearly 
 sixfold (${4}\choose{2}$) degenerate.
The spectrum in the absence of the spin-orbit interaction 
is displayed in Fig. \ref{ep}(a) with the
dominant contributions to the two-electron wave function
given in the Figure near the energy levels. 
At $B=0$, one finds a ground-state triplet, and an excited
state doublet that is next followed by a singlet. 

Let us discuss the structure of the energy levels in Fig. \ref{ep}(a).
For the spin-down polarized ground-state $K'_d K_d$ \cite{uwaga}
the interaction integral $I=\langle \psi|\frac{1}{r_{12}}|\psi\rangle $ with the antisymmetrized wave functions 
is $I =\langle K'_d (1) K_d(2) | \frac{1}{r_{12}} |K'_d (1) K_d(2)  \rangle-\langle K'_d (1) K_d(2) | \frac{1}{r_{12}} |K_d (1) K'_d(2)  \rangle\equiv E_C-E_{X}$, where $E_C$ and $E_{X}$ are the 
Coulomb and  exchange integrals, respectively. The exchange integral
 is non-zero due to intervalley scattering induced by the short-range component of the electron-electron interaction potential. The same result is obtained for the other
spin-polarized state of the ground-state triplet $K'_uK_u$. 

For the $K'$-valley-polarized state $K'_d K'_u$ the 
interaction integral is 
$I_2=E_C \mp \langle K'_d (1) K'_u(2) | \frac{1}{r_{12}} |K'_u (1) K'_d(2)  \rangle=E_C \mp 0 $. 
Here, the exchange integral vanishes due to the spin mismatch
of the wave functions for both the first and the second electron in the integral.
 The same result is obtained
for the $K$-valley-polarized state $K_dK_u$ state.
Therefore, at $B=0$ the valley-polarized states $K'_dK'_u$ and $K_dK_u$ form a degenerate doublet at the energy $E_x$ above the ground-state energy level [see Fig. \ref{ep}(a)].  

In Fig. \ref{ep}(a) there are two energy levels 
which are neither spin nor valley polarized [($K'_dK_u\pm K'_uK_d)/2^{1/2}$]. For these states 
the interaction integral is $I_3=C\mp E_X$.
Hence, the lower-energy state 
enters the ground-state
triplet, and the other is the singlet at the energy of $2E_X$ above the ground state.

The role of the short-range component of the Coulomb interaction for the  intervalley exchange
can be illustrated by comparison of Fig. \ref{ep}(a) and Fig. \ref{ep}(b).
In Fig. \ref{ep}(b) we plotted  the  spectrum that is obtained
for the Coulomb potential $1/r_{12}$ replaced
by a function $V(r_{12})=\min (\frac{1}{r_{12}},\frac{1}{3a})$ , where $a$ is the silicene lattice constant. 
This potential removes the shortest-range maximum of the Coulomb interaction. 
In Fig. \ref{ep}(b) we notice that the splitting of the energy levels is reduced nearly 10 times.  
The modified Coulomb interaction is used only in Fig. \ref{ep}(b).

Figure \ref{ep}(c) shows the results with the spin-orbit
interactions included. The  states that were not polarized neither in spin nor in valley 
that in Fig. \ref{ep}(a) were split only by a small value of $2E_X$
now differ in the energy much more, i.e., by twice 
 the single-electron spin-orbit splitting  ($2\Delta_{SO}$). 

In the ground-state of Fig. \ref{ep}(c) the configuration $K'_dKu$ with both electrons in the states
of the lowest single-electron doublet [cf. Fig. \ref{widma1eb}(a,b)] is dominant.  The contribution of the excited doublet with interchanged valley indices $K'_uK_d$ is only $\simeq 4$\%. This contribution corresponds to both electrons in the higher-energy single-electron doublet. Since the contribution is small, the intervalley exchange energy  
 is negligible.

In the  four states in the center 
of the spectrum in Fig. \ref{ep}(c) one of the electrons occupies a single-electron state of the lower doublet,
and the other a state of the higher doublet. 
These states at $B=0$ form two doublets. The energy of the spin-polarized doublet
is lower by the intervalley exchange $E_X$ as in Fig. \ref{ep}(a).

\subsection{Transitions in the alternate electric field: the single-electron}

For the system in the alternate electric field the transition times are proportional 
to the inverse of the  dipole matrix elements which are displayed by lines in Fig. \ref{tramael}(a) as  functions of the potential depth $w$.
The results are calculated for the $K'_d$ ground
state as the initial state. In Fig. \ref{tramael}
the spin-flip (valley-flip) transition is obtained for $K'_u$ ($K_d$) as the final state. The matrix element
for the simultaneous flip of both the valley and the spin 
$K'_d\rightarrow K_u$ is too small to fit in the Figure. 
We find that the matrix element for the valley flip
vanishes for large $w$. The armchair edge of the flake
is responsible for the intervalley coupling \cite{zarenia}. For
large $w$ the confined states are entirely localized
within the dot (Fig. \ref{wiazanie}), and the intervalley flip by the electric
field is no longer possible. 

In Fig. \ref{tramael}(b) we translate the matrix elements
to the transition times as obtained for the amplitude 
of the electric field of $F_{AC} =2$kV/cm.  
For shallow confinement ($w\simeq 150$ meV) the valley flips are very fast, of the order of 100 ps. For $w=400$ meV the valley flip times are as large as 20 $\mu$s. 
On the other hand the spin-flip times
are about 1 to 2 ns and weakly depend on the potential depth. 
Note, that the spin flip time slightly increases at large $w$, which
might be counterintuitive since the electric field and thus the extrinsic contribution to the Rashba coupling is 
enhanced for larger $w$. However, the orbital extent of the wave functions decreases with growing $w$, and the latter effect is dominant for the values of the dipole matrix elements.

Figure \ref{czas1} shows the results of the time-dependent
calculation as function of the AC frequency $\nu$
for $F_{AC}=2$ kV/cm with values of $w$.
The magnetic field is set to 1 T, the system is started in the $K'_d$ ground state
and the simulation lasts 3.74 ns. 
The plots in Fig. \ref{czas1} show the maximal square
of the  projection of the time dependent wave function
on the stationary states. For both the spin transition 
to $K'_u$ and the valley transition
to $K_d$  we list the times
upon which occupation of the final state exceeds 10\% 
and 50\%.  For $w=250$ meV [Fig. \ref{czas1}(c)] the valley transition time is already much longer than time covered by the simulation.

\subsection{Transitions in the AC electric field: the electron pair} 

The dot symbols in Fig. \ref{tramael} indicate the transition
matrix elements and transition times for the electron pair.
The results were calculated for the ground-state
of Fig. \ref{ep}(c) to $K'_dK'_u$ or $K'_d K_d$ excited states,
where the listed spin-valley configurations stand for the two-electron Slater determinants.
The dominant contributions to the  ground state wave function is $\psi=(K'_dK_u+cK'_uK_d)/N$, where $N$
is the normalization constant and $c$ is small  $(|c|^2 <<1)$. 
The  transition to the $K'_dK'_u$  state is 
identified with the valley transition since it
involves the valley flip
$K'_dK_u\rightarrow K'_d K'_u$ in the dominant term of the 
ground-state wave function, or $K'_uK_d\rightarrow K'_uK'_d$ in 
the smaller term. 
For the similar reason the transition from the ground state to $K'_d K_d$ can be identified with the spin flip. 

Figure \ref{tramael} shows that the spin-flip times as calculated for the electron pair are very close to  the results for the single-electron. The two-electron results for the valley flip
follow the single-electron trend, but   for large $w$,
when the system is separated from the  edge,
the short range component of the electron-electron enhances the valley transition rates.

\begin{figure}
\begin{tabular}{l}
(a) \includegraphics[width=0.9\columnwidth]{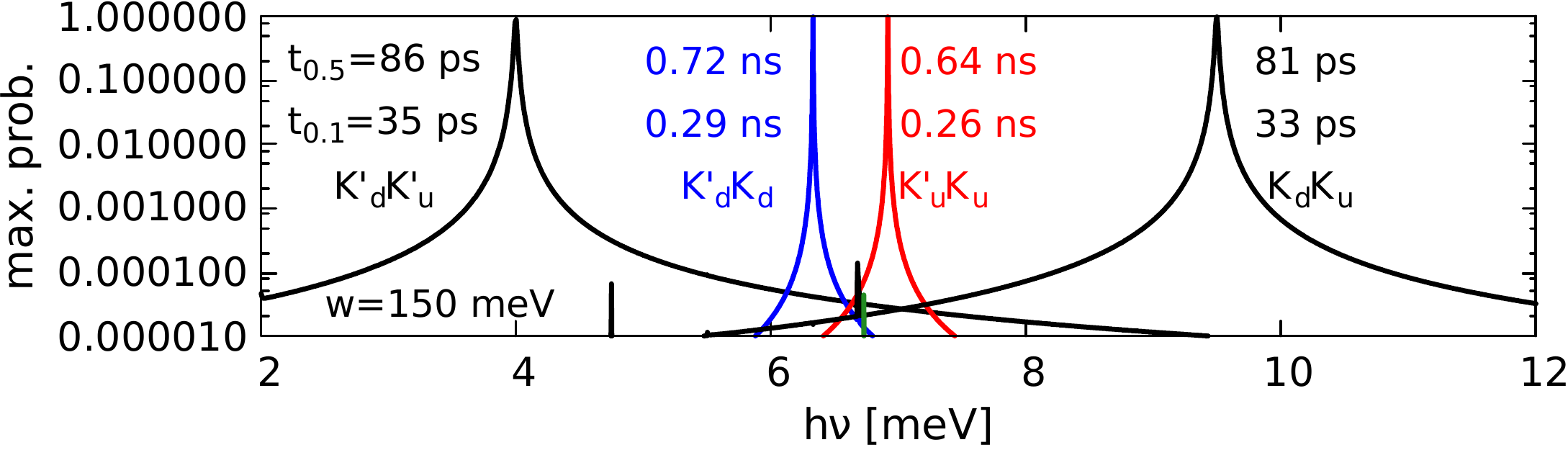} \\
(b) \includegraphics[width=0.9\columnwidth]{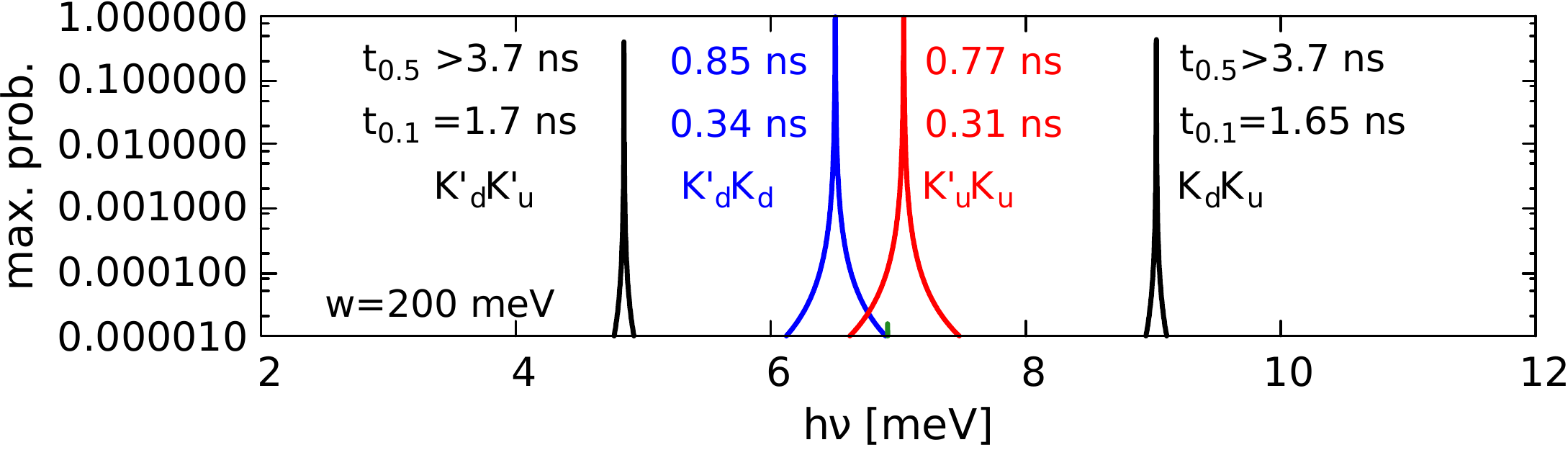}   \\
(c) \includegraphics[width=0.9\columnwidth]{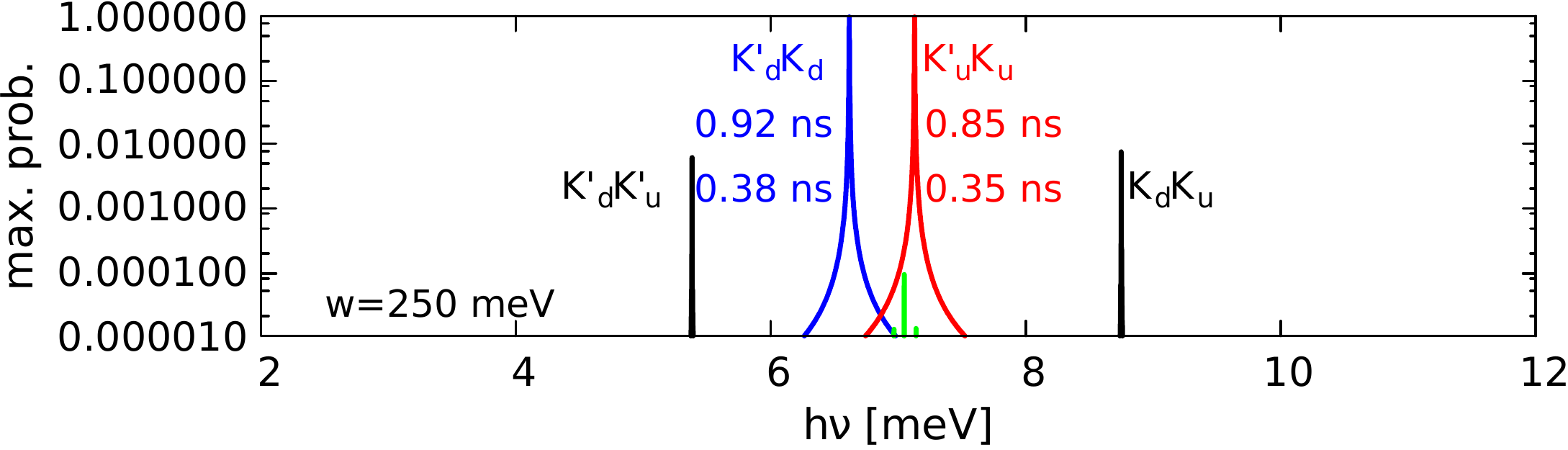}  \\
(d) \includegraphics[width=0.9\columnwidth]{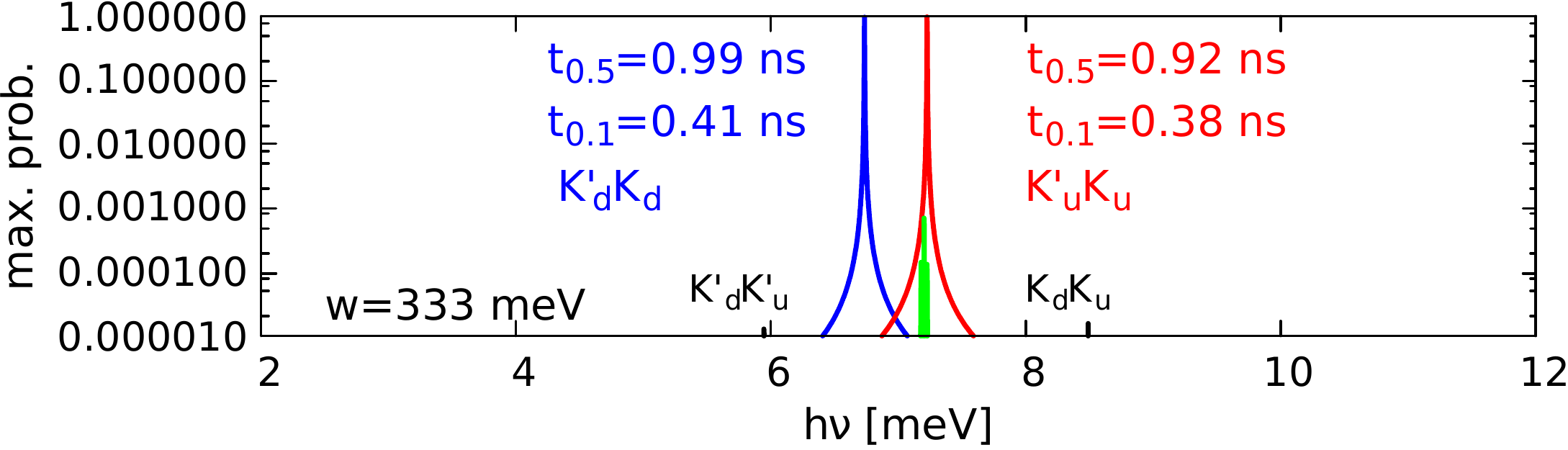}  \\
\end{tabular}
\caption{Counterpart of Fig. \ref{czas1} for two excess electrons within the QD. In the initial condition
the two electrons are set in the ground-state for 1T
[for the spectrum see Fig. 4(a)]. The transition to the valley polarized states (black lines) correspond to the valley flip of one of the electrons. The transition to the spin-polarized states (blue and red lines) require spin flip for one of the electrons in the ground state. The green line
shows the maximal occupation probability for the highest energy level of Fig. 4(a). The direct transition to these energy level is forbidden, since the matrix element is zero. A low peak for occupation of this state appears at half the energy difference and has a character of a two-photon transition. }
 \label{czas2}
\end{figure}

The results for the time-resolved simulation of the driven transitions in the two-electron quantum dot are presented in 
Fig. \ref{czas2} for the ground state in the initial condition [cf. Fig. \ref{ep}(c)].
The transitions to spin-polarized excited state $K'_dK_d$  and $K'_uK_u$ both involve the spin flip of one of the electrons in the ground state.
At resonant energies it takes the driven wave function about 0.3 ns (1 ns)   to reach about  10\% (50\%) admixture of the spin-polarized state.
The valley flips which are  very fast (several ps) for shallow confinement potential,  dissappear for larger $w$.

\begin{figure*}
\begin{tabular}{l}
\begin{tabular}{lll}
(a) \includegraphics[width=0.45\columnwidth]{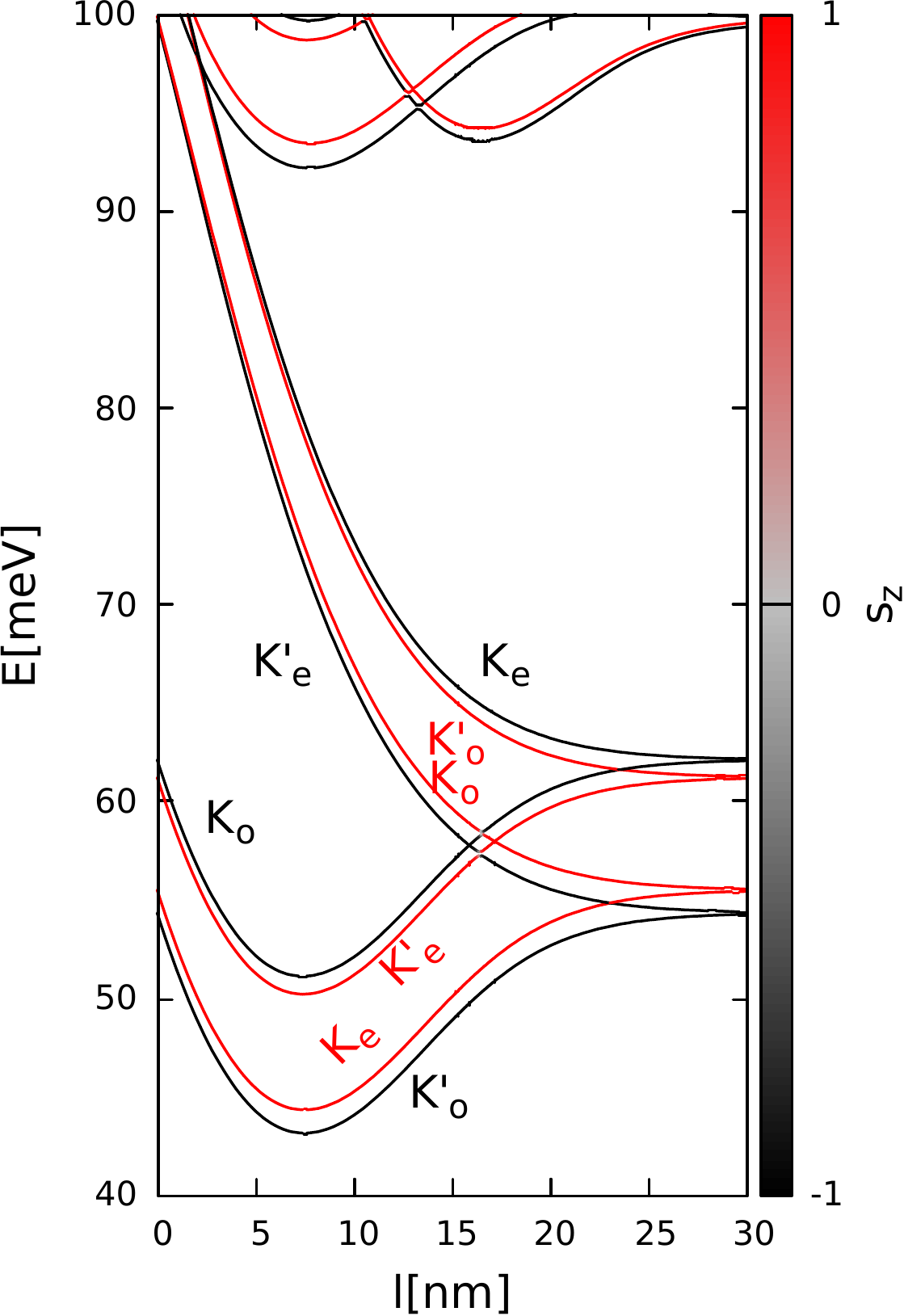} &
(b) \includegraphics[width=0.45\columnwidth]{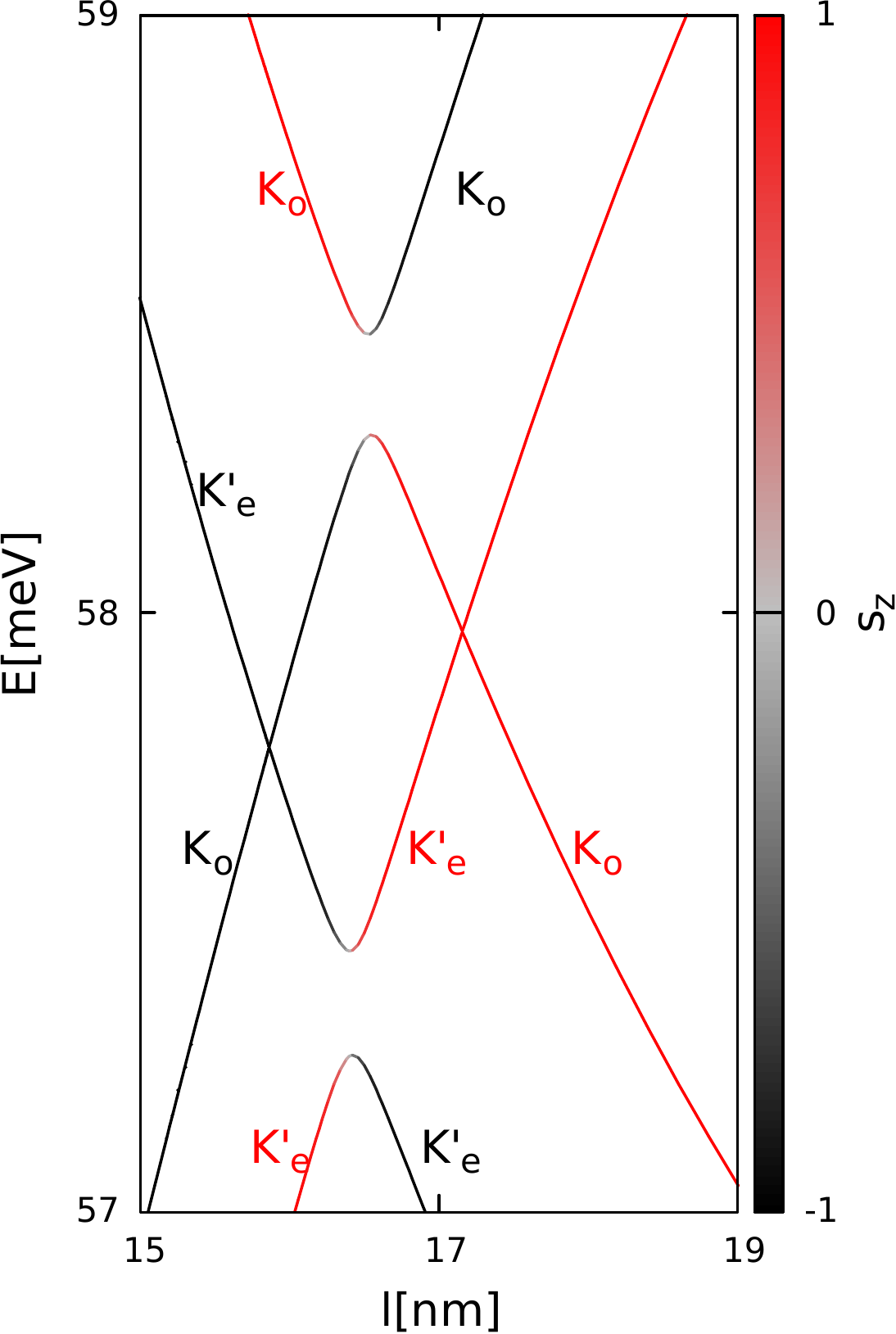} &
(c) \includegraphics[width=0.45\columnwidth]{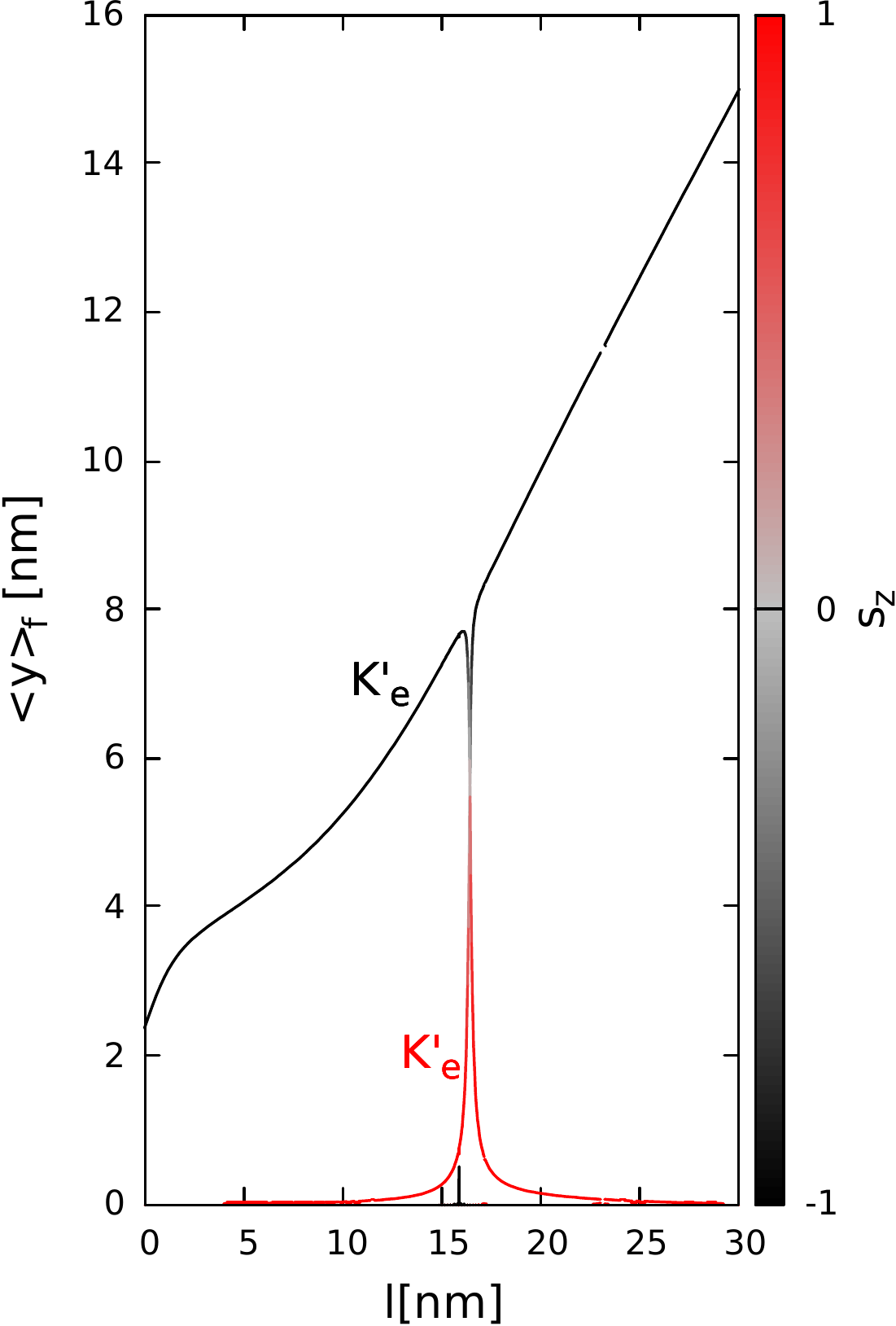}    
\end{tabular} 
\\ 
\begin{tabular}{llll}
(d) \includegraphics[width=0.45\columnwidth]{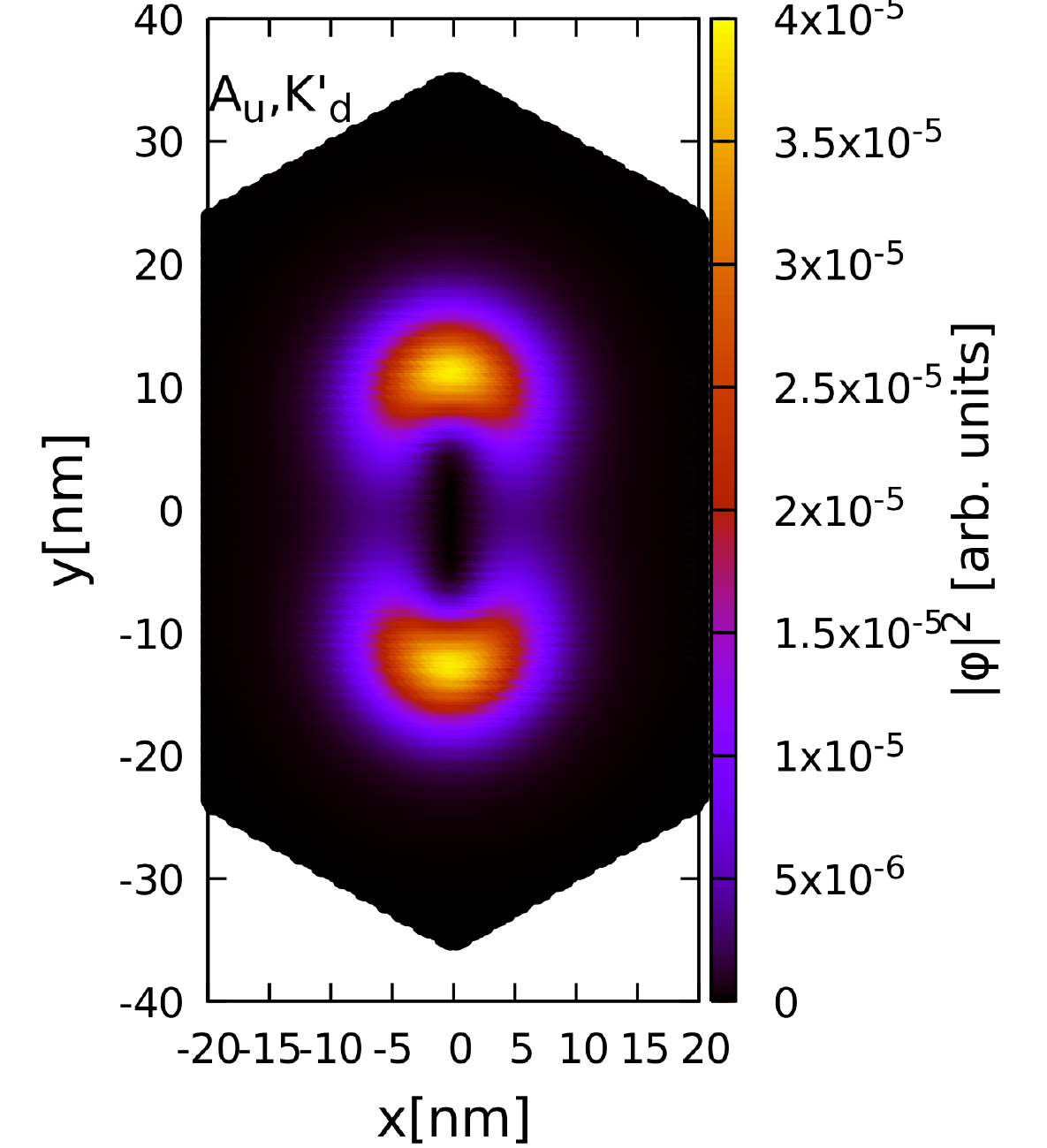} &
(e) \includegraphics[width=0.45\columnwidth]{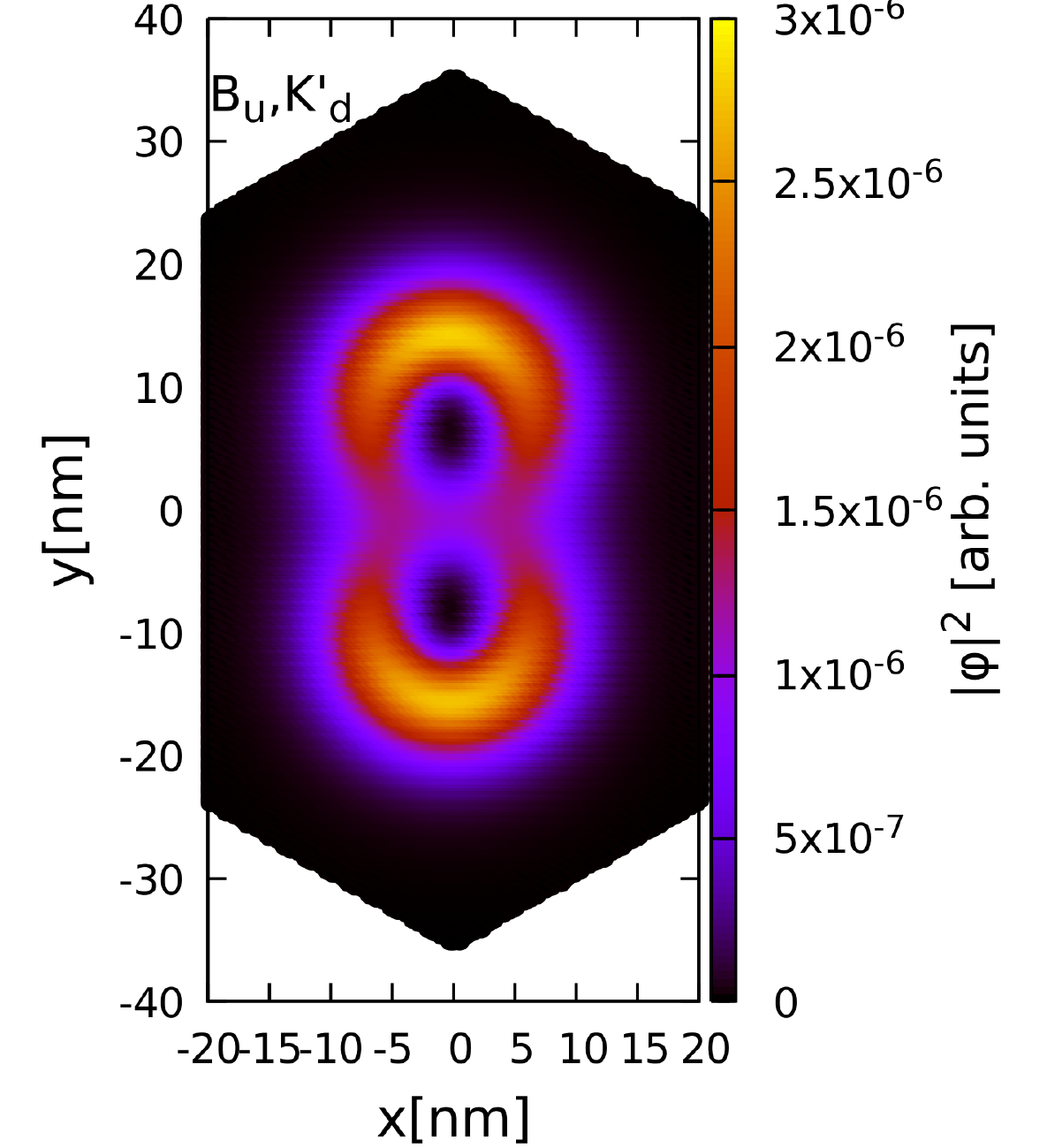} &
(f) \includegraphics[width=0.45\columnwidth]{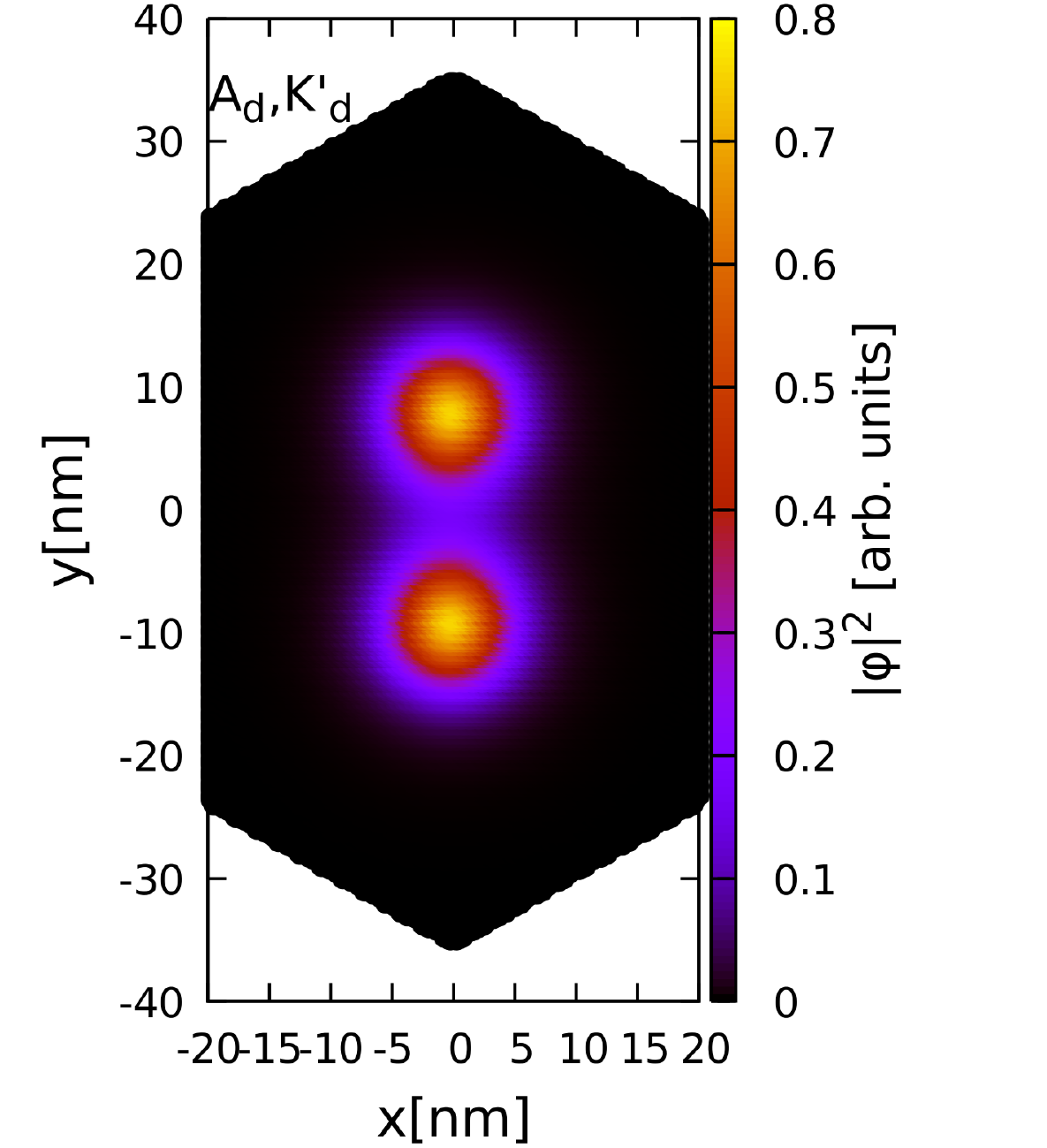}   & 
(g) \includegraphics[width=0.45\columnwidth]{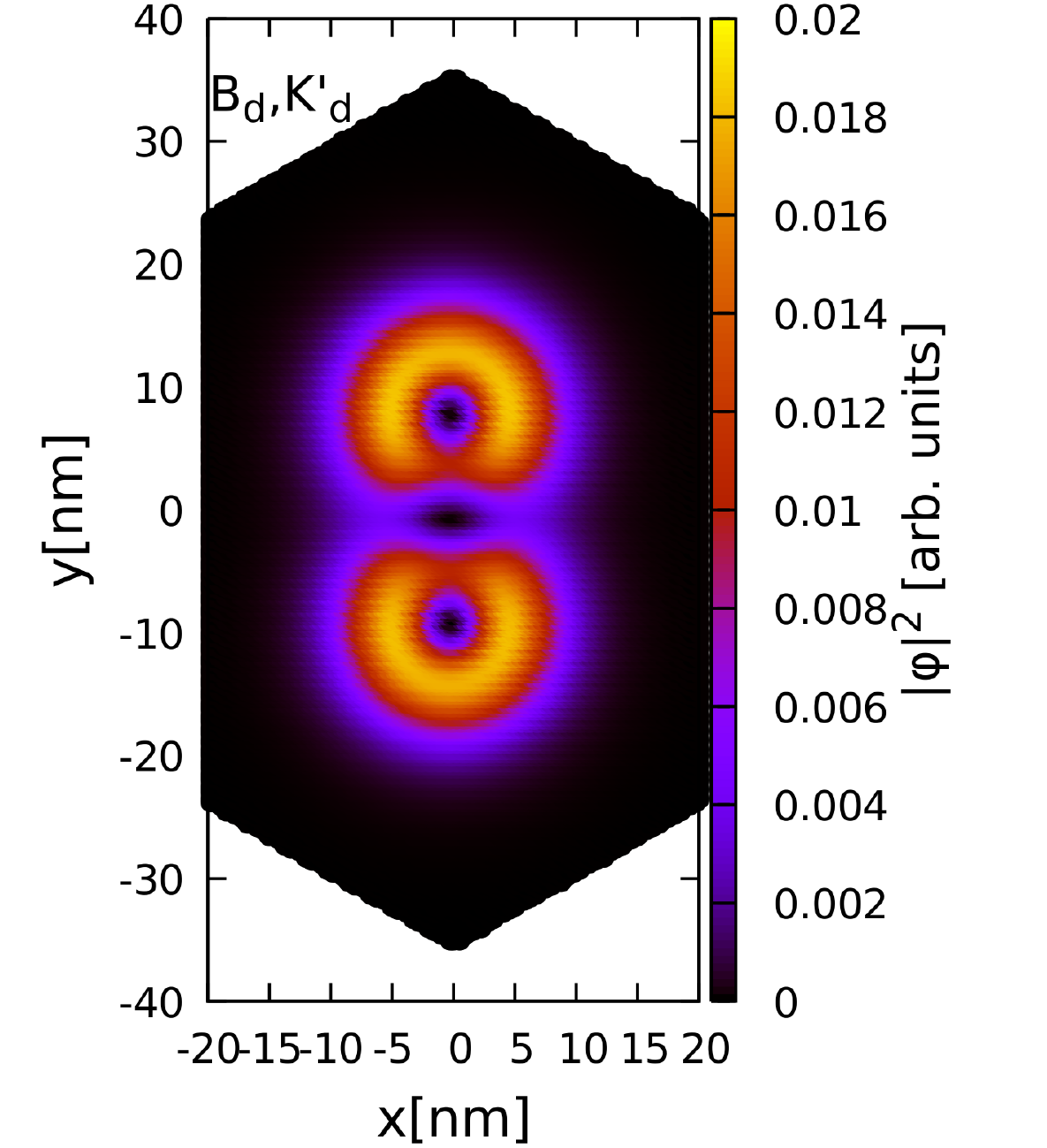}    \\
\end{tabular}
\end{tabular}
\caption{A single electron in a pair of identical dots with $w_1=w_2=200$ meV for $B_z=0.5$ T. (a) The energy spectrum as a function
of the distance between the dot centers. The color indicates
 the z-component of the spin. The subscript $e$ or $o$ stands for the states of even and odd parity (see text). 
(b) Zoom of the avoided crossings between the energy levels
of opposite spin.  
(c) The matrix element for the transition from the $K'_o$ ground state to the excited states. At this scale only
the valley conserving transitions are resolved. Moreover,
the direct transitions to $K'_o$ driven by the AC field oriented in the $y$ direction are forbidden by the parity symmetry.
(d-f) The probability density for the $K'_o$ ground-state, of odd generalized symmetry, with spin-down orientation and $K'$ valley for $B=0.5$ T and $l=18$ nm. 
Panels (d-g)  correspond to spin orientations and sublattices in the following order (from left to right):
 $A\uparrow$, $B\uparrow$, $A\downarrow$, $B\downarrow$. The corresponding components are odd, even, even and odd,
spatial parity with respect to the point inversion with respect to the center of the system.
 }
 \label{d1k}
\end{figure*}

\begin{figure}
\begin{tabular}{l}
(a) \includegraphics[width=0.52\columnwidth]{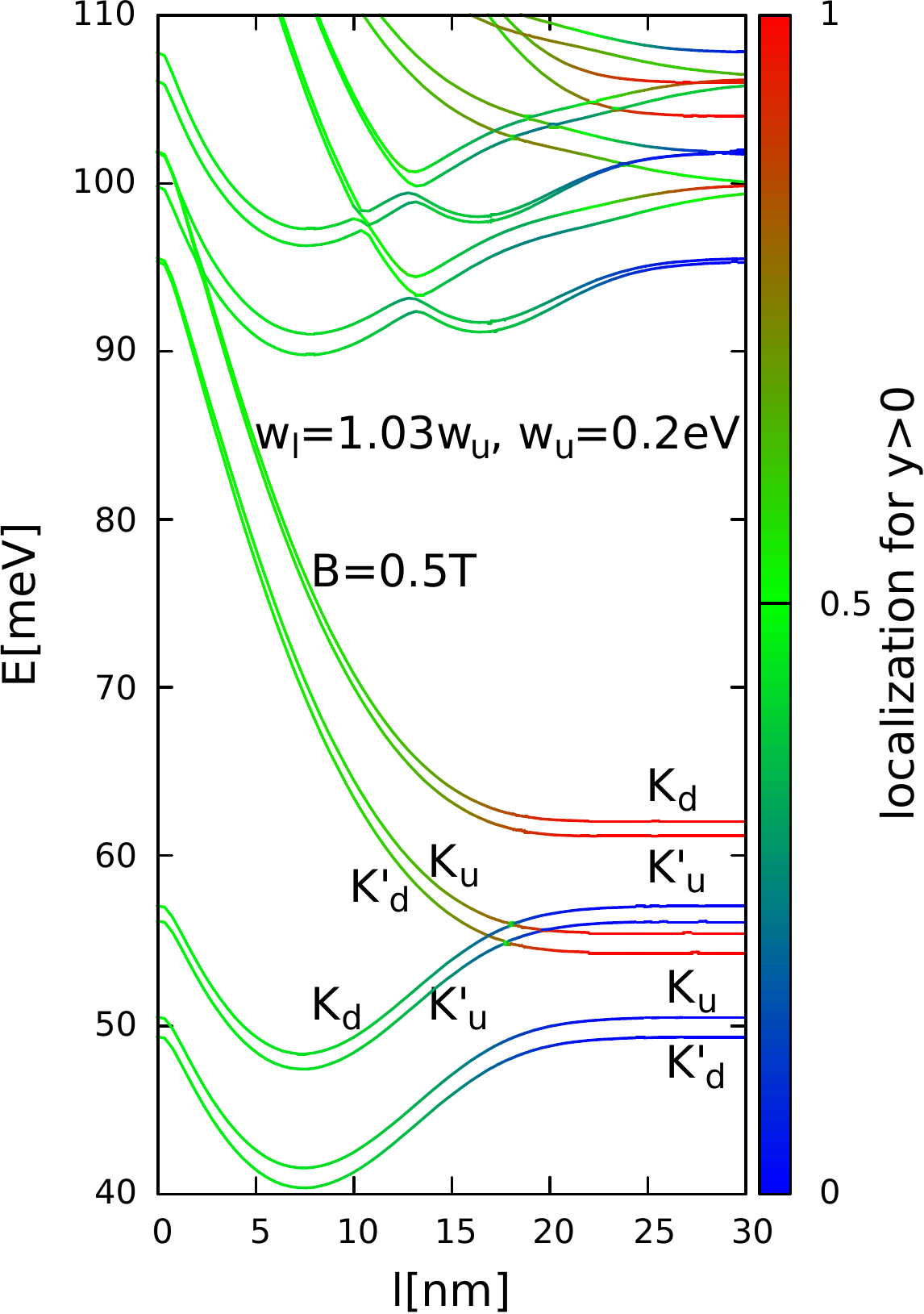} \\
(b) \includegraphics[width=0.52\columnwidth]{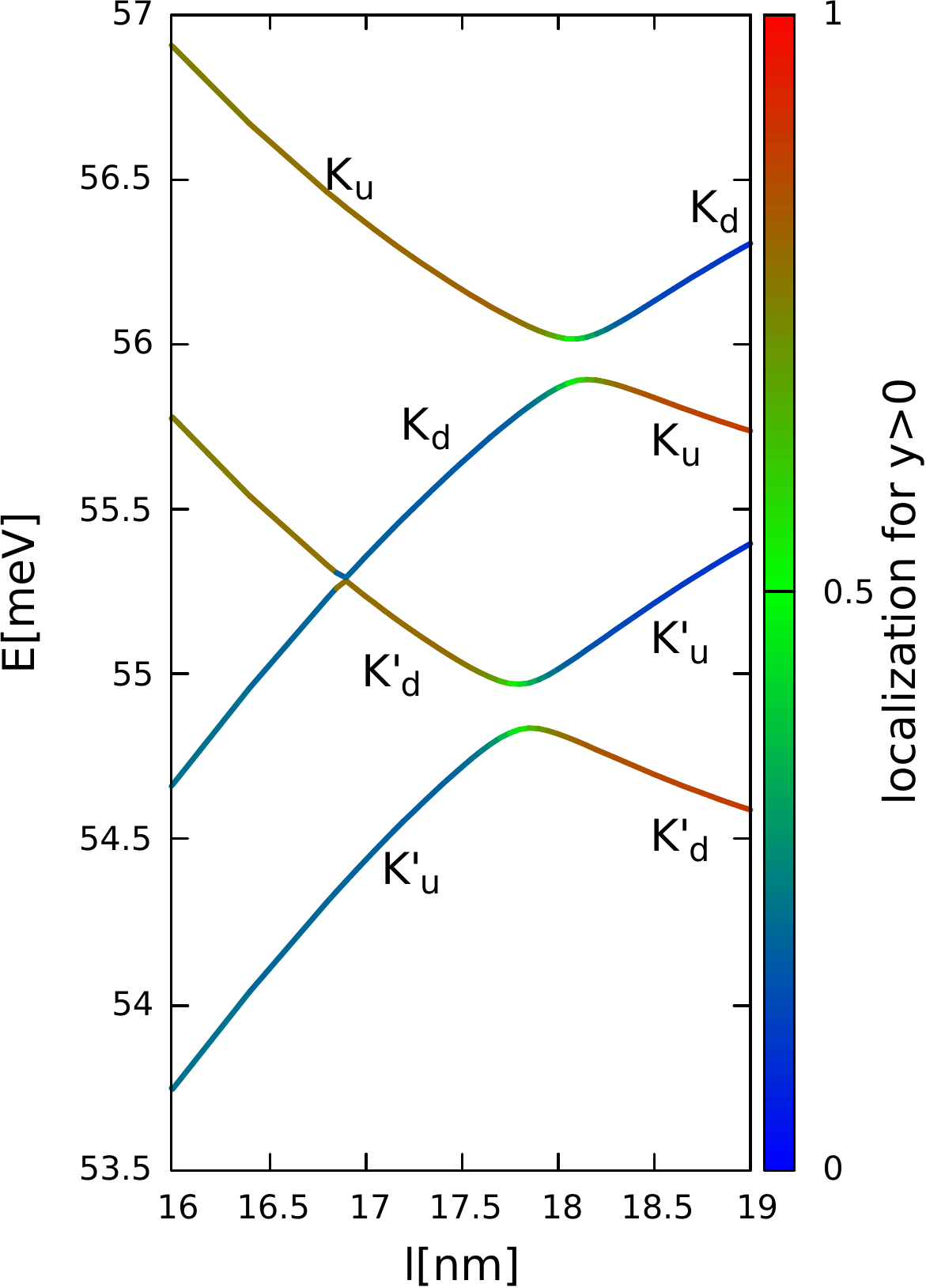}   \\
(c) \includegraphics[width=0.45\columnwidth]{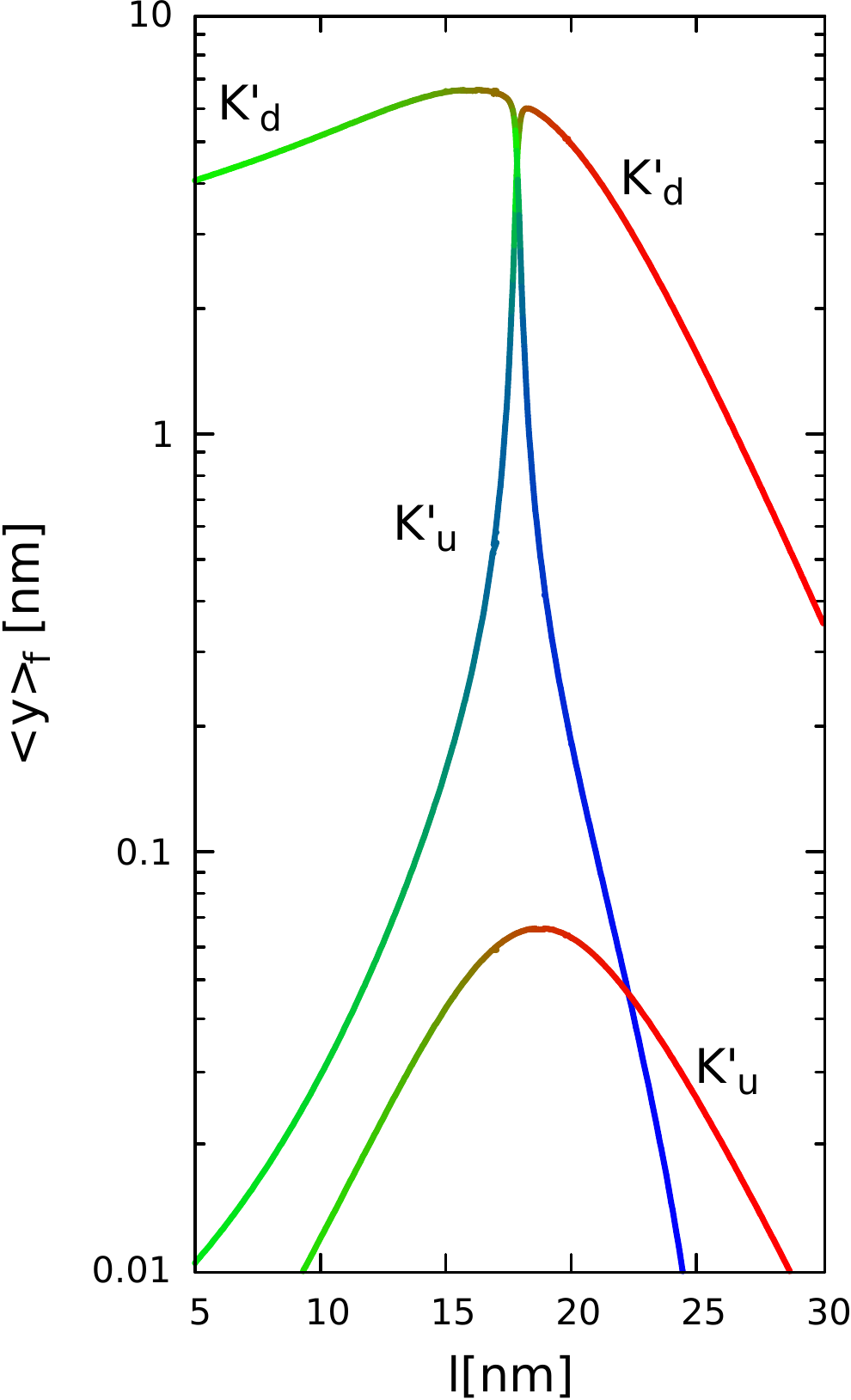}   \\
\end{tabular}
\caption{(a) Energy spectrum for a single electron in a
weakly asymmetric
double dot system, with $w_l=1.03w_u$ and $w_u=200$ meV and  $B=0.5$ T as a function of the interdot distance $l$.
 The color of the line indicates the localization
in the lower dot (blue) or upper dot (red). 
(b) The zoom on the avoided crossing open by the Rashba interaction. 
(c) Transition matrix elements from the $K'_d$ ground state
to the excited states. The color of the line indicates 
the localization of the excited state.}
 \label{wasd}
\end{figure}

\begin{figure}
\begin{tabular}{l}
(a) \includegraphics[width=0.75\columnwidth]{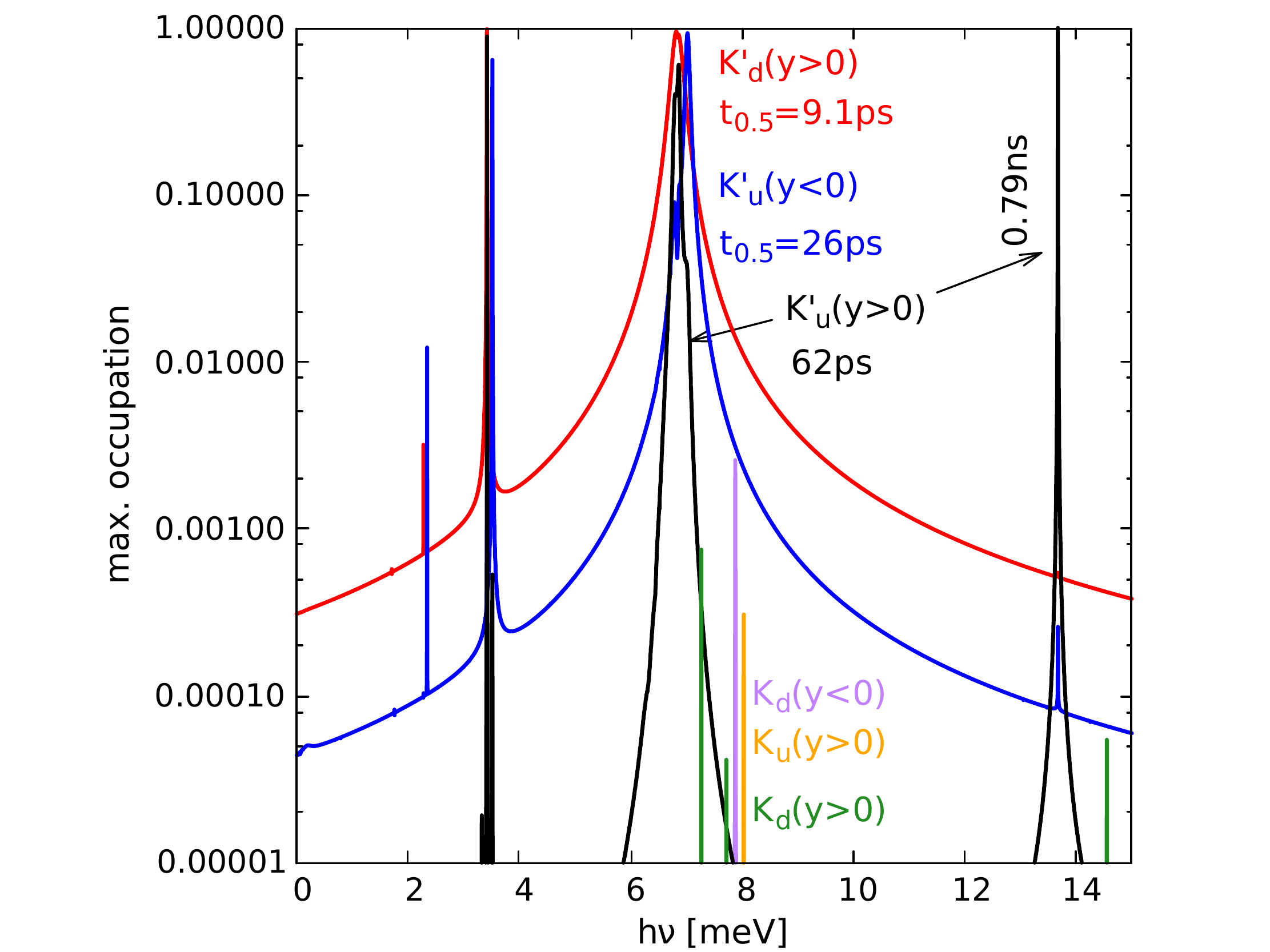} \\
(b) \includegraphics[width=0.75\columnwidth]{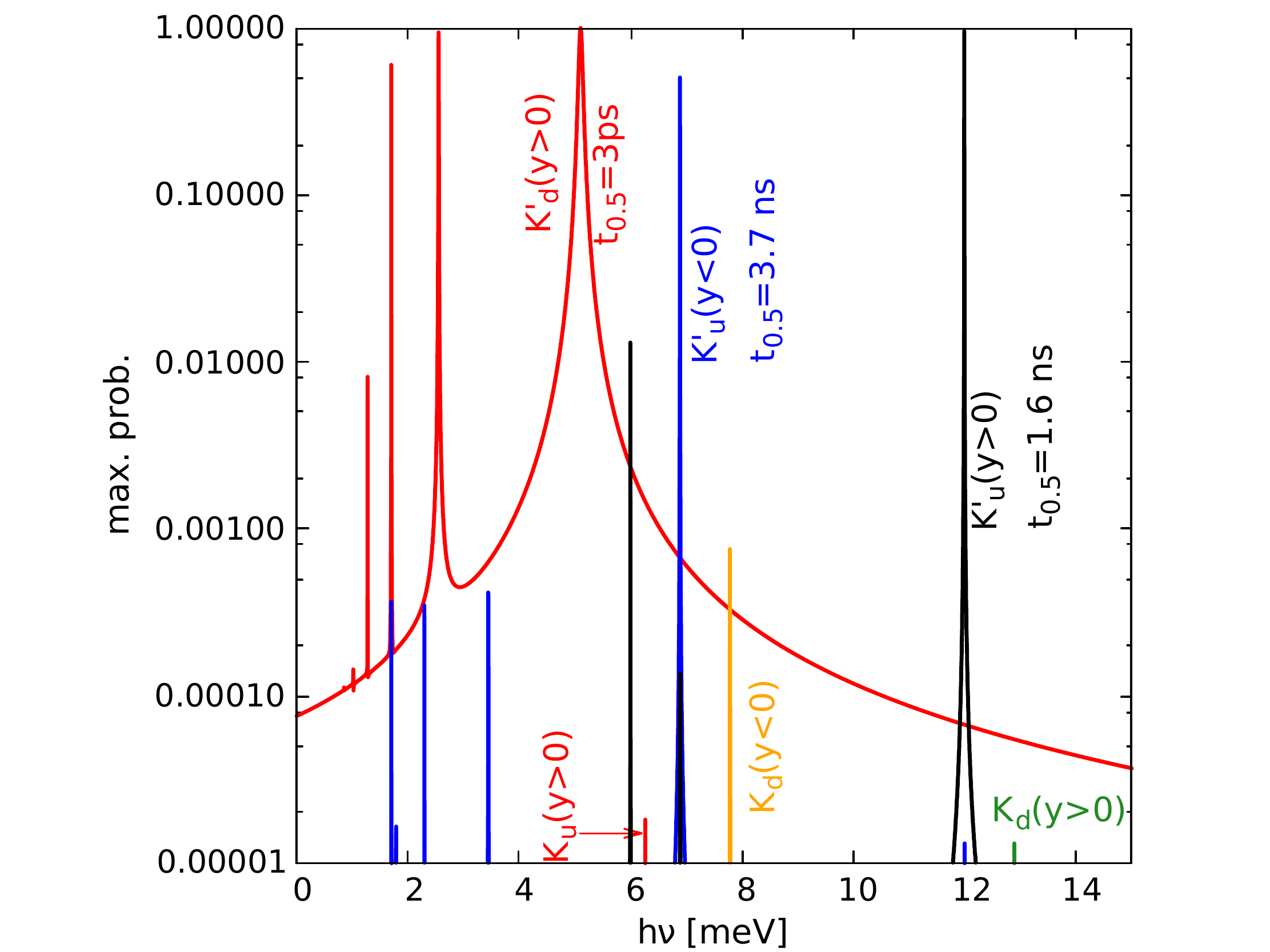} \\
\end{tabular}
\caption{
Transitions in the alternate electric field for weakly  
asymmetric double dot of Fig. \ref{wasd}, $B=0.5$ T and $F_{AC}=0.2$ kV/cm. The results present the maximal
occupation probability for the simulation time of 3.74 ns. 
The system is initially in the $K'_d$ ground state localized
mostly in the lower quantum dot which is deeper. 
In the plot we provide the time required for a given excited state to appear with a 50\% contribution to the time dependent wave function. 
Results for  $l=18$ nm (a) and $l=24$ nm (b).}
 \label{czas119}
\end{figure}

\begin{figure}
\begin{tabular}{l}
(a) \includegraphics[width=0.65\columnwidth]{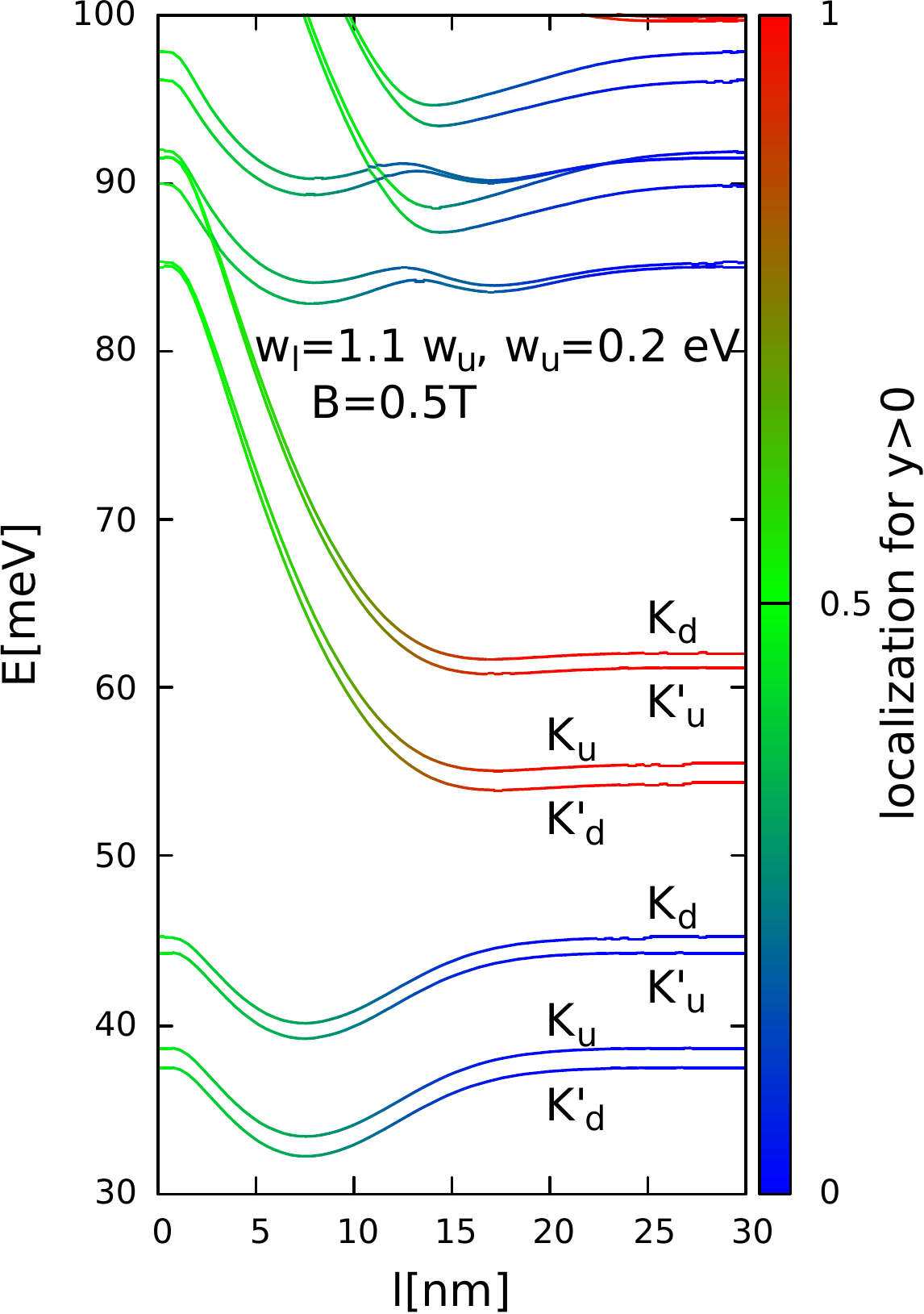} \\
(b) \includegraphics[width=0.65\columnwidth]{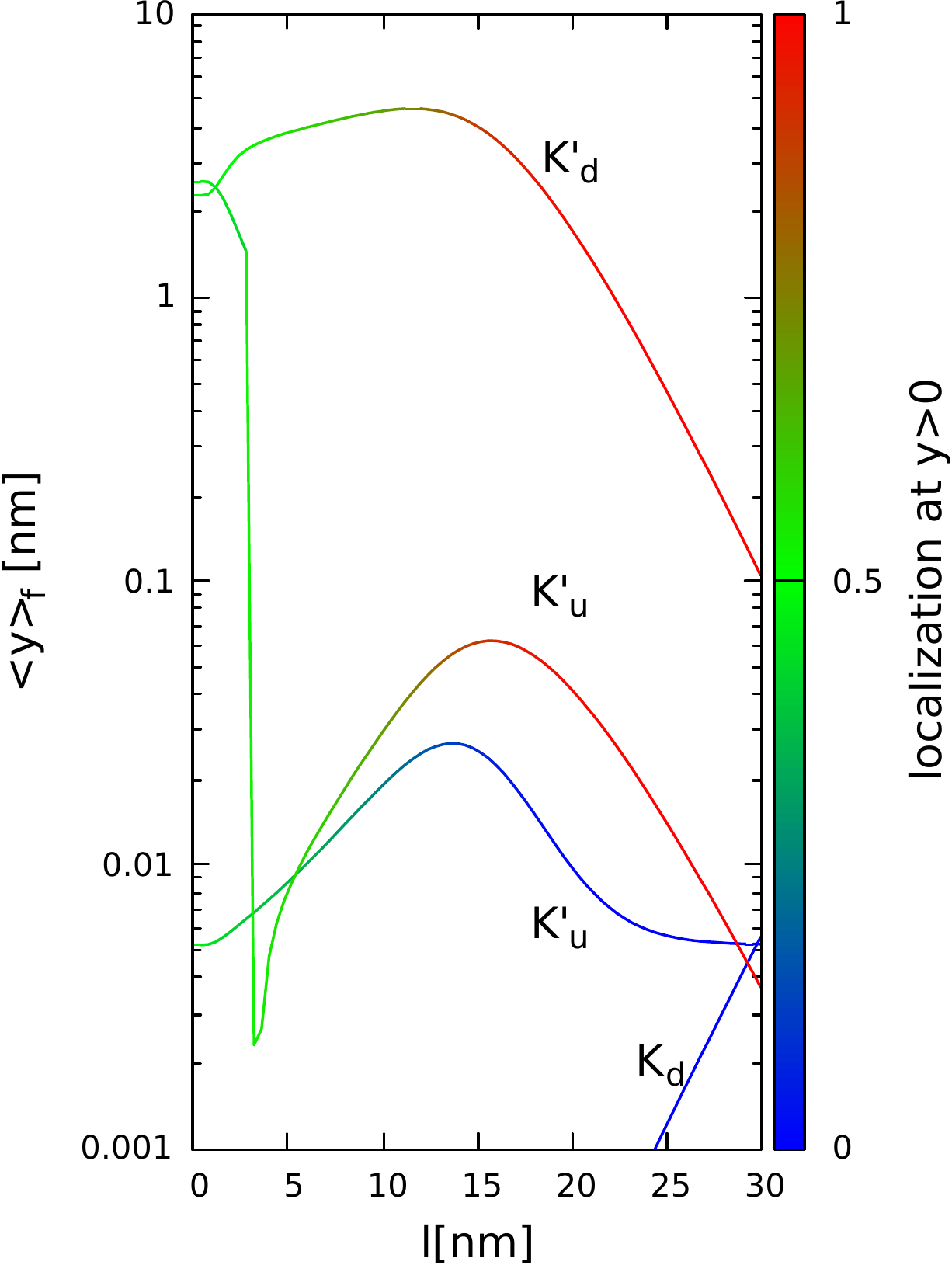}   \\
\end{tabular}
\caption{
(a) Energy spectrum for a single electron in an asymmetric
double dot system, with $w_l=1.1w_u$ and $w_u=200$ meV, for $B=0.5$ T as a function of the interdot distance $l$.
 The color of the line indicates the localization
in the lower dot (blue) or upper dot (red). 
(b) Transition matrix elements from the $K'_d$ ground state
to the excited states. The color of the lines indicates 
the localization of the excited state. } 
 \label{asy11}
\end{figure}

\begin{figure}
\begin{tabular}{l}
(a) \includegraphics[width=0.75\columnwidth]{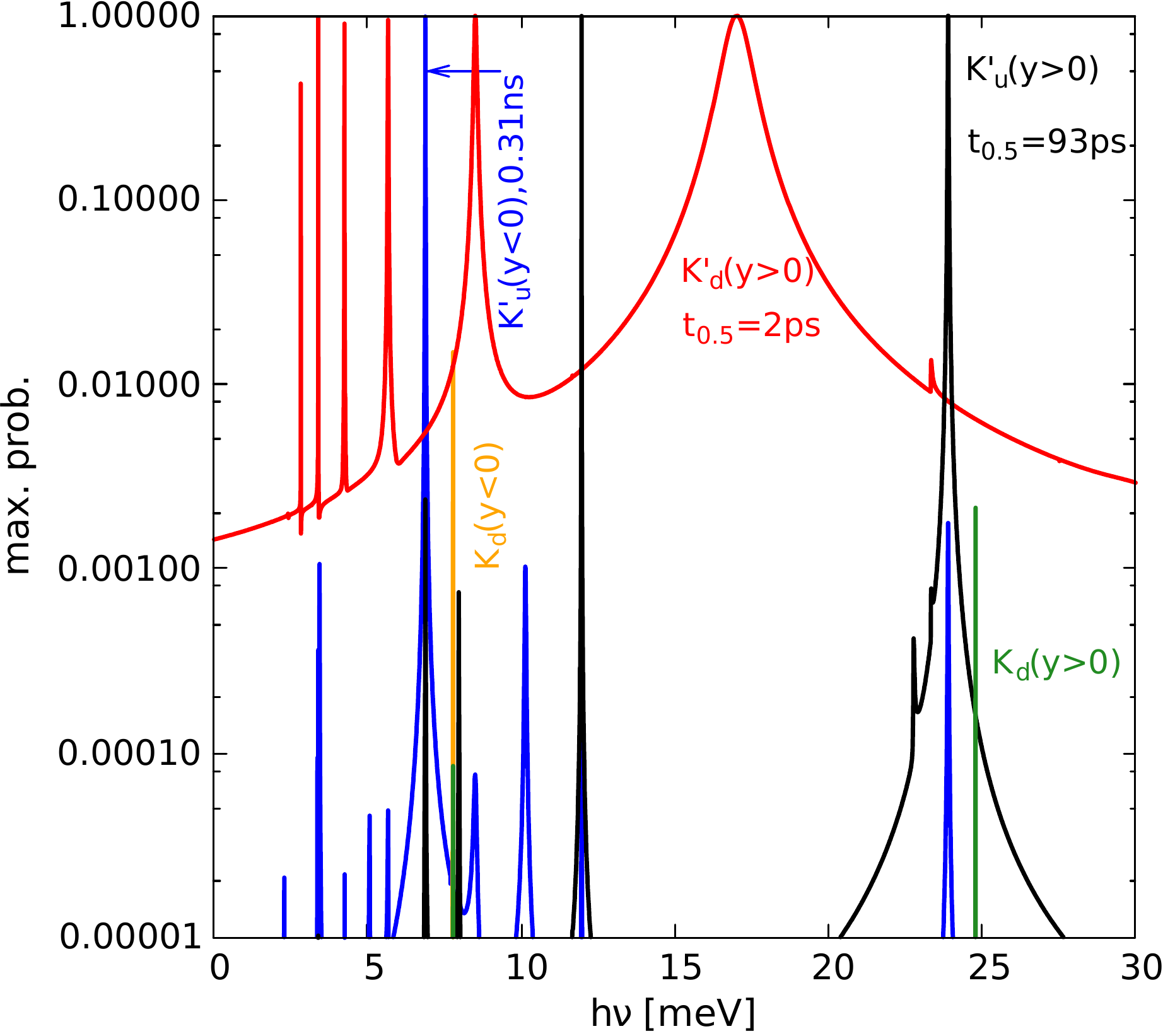} \\
(b) \includegraphics[width=0.75\columnwidth]{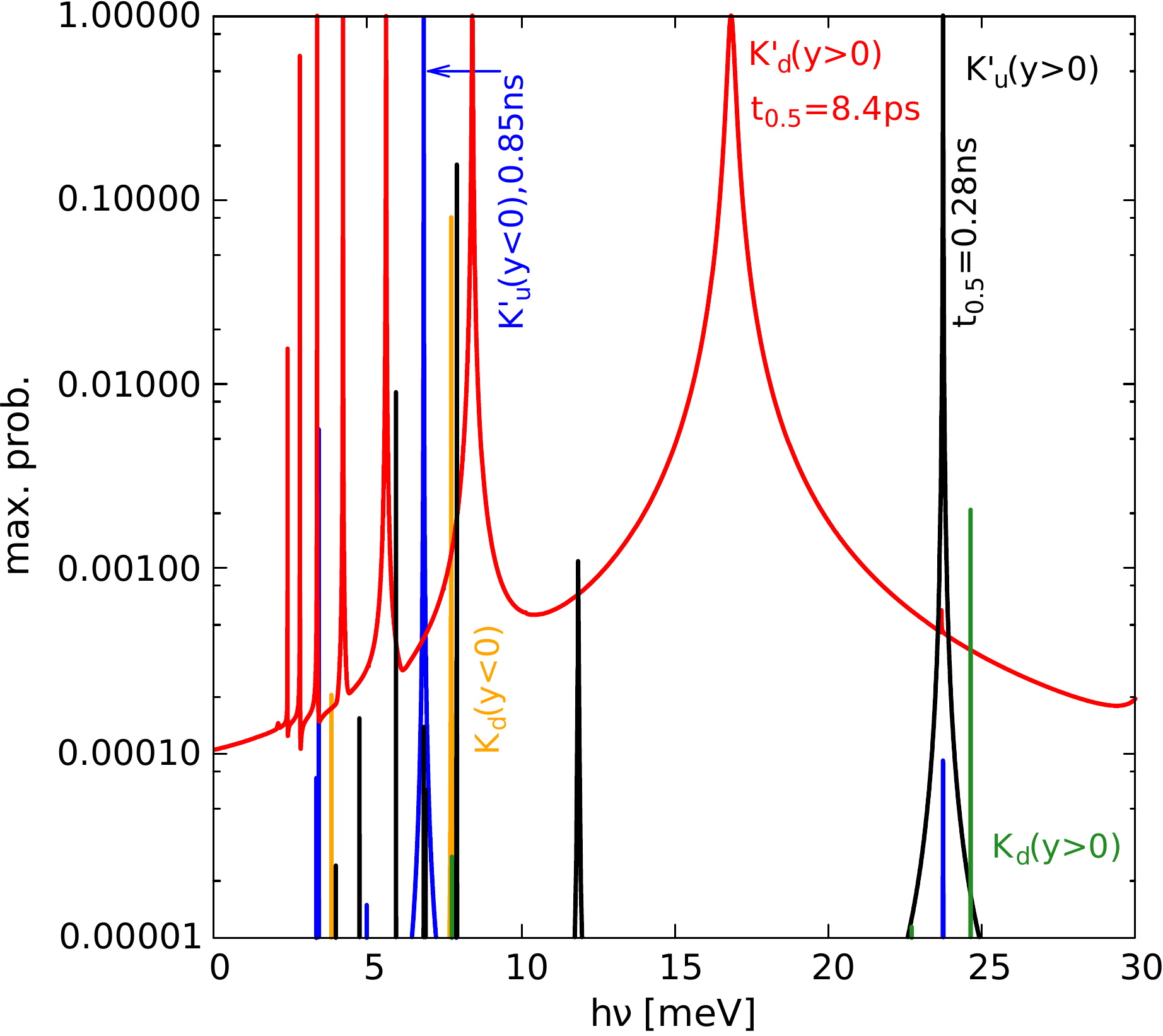} \\
\end{tabular}
\caption{
Transitions driven by AC electric fields for 
asymmetric double dot of Fig. \ref{asy11}, the magnetic field of 0.5 T and the amplitude of 
the AC field of 2 kV/cm. The results present the maximal
occupation probability for the simulation time of 3.74 ns. 
The system is initially in the $K'_d$ ground state localized
mostly in the lower quantum dot (the deeper one). 
We give the time required for a given excited state to appear with a 50\% contribution to the time dependent wave function. 
Results for  $l=18$ nm (a) and $l=24$ nm (b).}
 \label{czasyasy11}
\end{figure}

\begin{figure*}
\begin{tabular}{lll}
(a) \includegraphics[width=0.65\columnwidth]{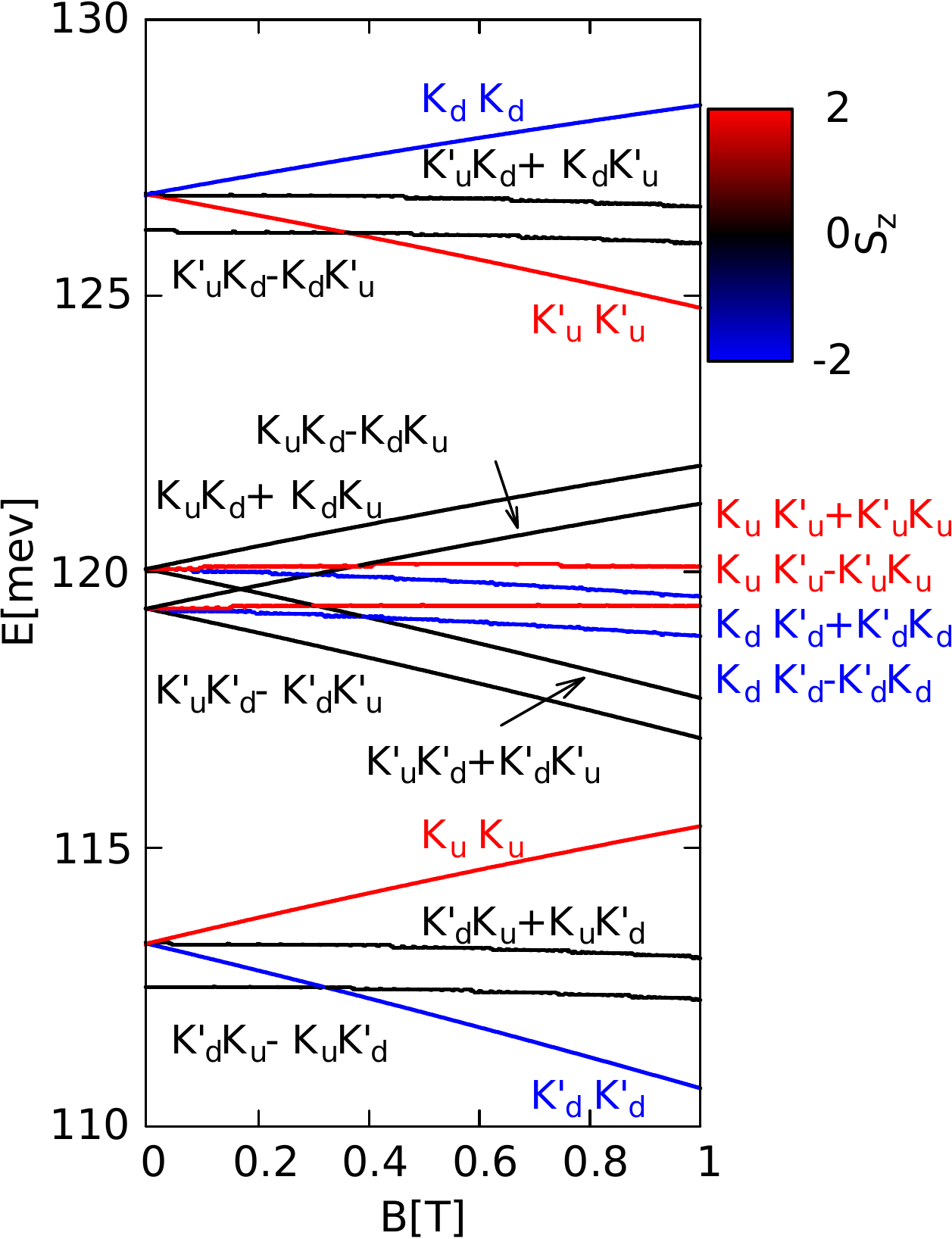} &
(b) \includegraphics[width=0.6\columnwidth]{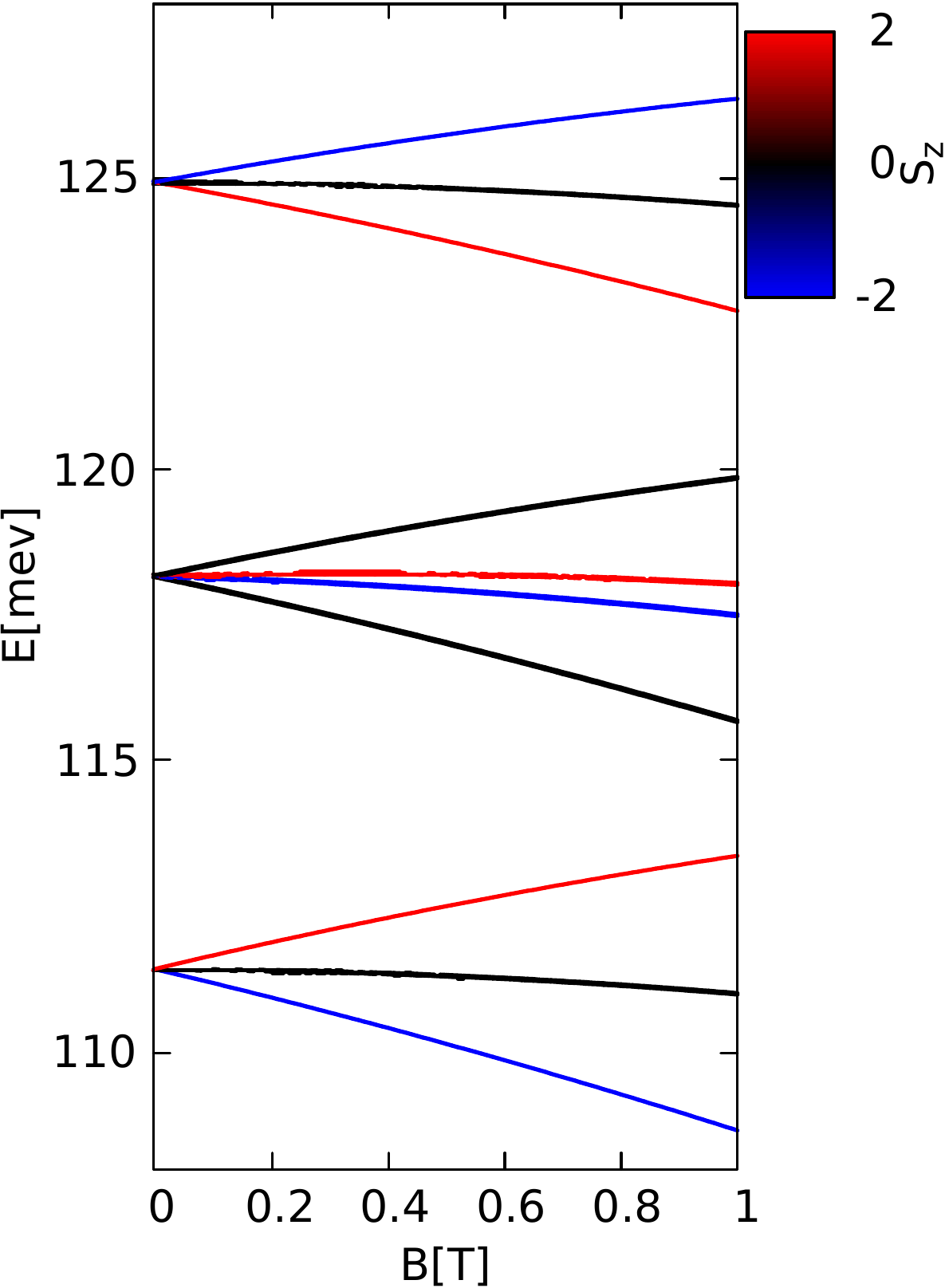} &
(c) \includegraphics[width=0.6\columnwidth]{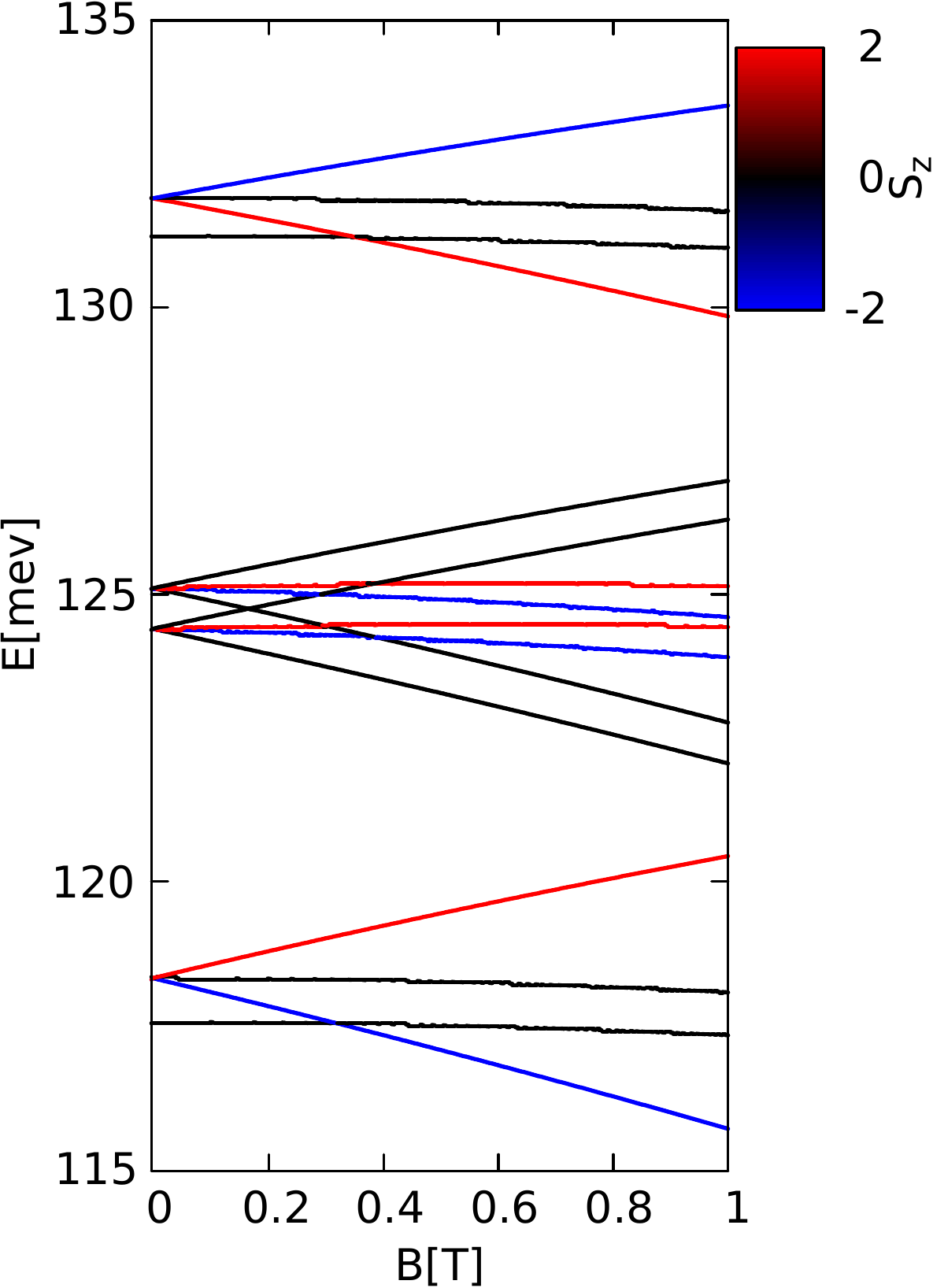} 
\end{tabular}
\caption{
Energy spectrum for two electrons confined in double quantum dots. In (a) and (b) an asymmetric
system with $w_l=1.03 w_u$ is considered  (cf. the single-electron results in Fig. \ref{wasd}).   The color of the line shows the $z$ component of the two-electron wave function in the $\hbar/2$ units.
In (a) the interdot distance of $l=18$ nm was applied. The energy levels that shift in pairs are split by the
interdot exchange energy that is due to tunnel coupling. 
For comparison in (b) $l=24$ nm was taken and the splitting can no longer be resolved. In (c) a symmetric system with $w_l=w_u$ is taken
with $l=18$ nm.  
In (a) the dominant contributions to the two-electron wave functions in the valley-spin space are given. Here, the left (right) term of each product is attributed to electron label 1 (2).
}
 \label{eper}
\end{figure*}

\begin{figure}

\includegraphics[width=1\columnwidth]{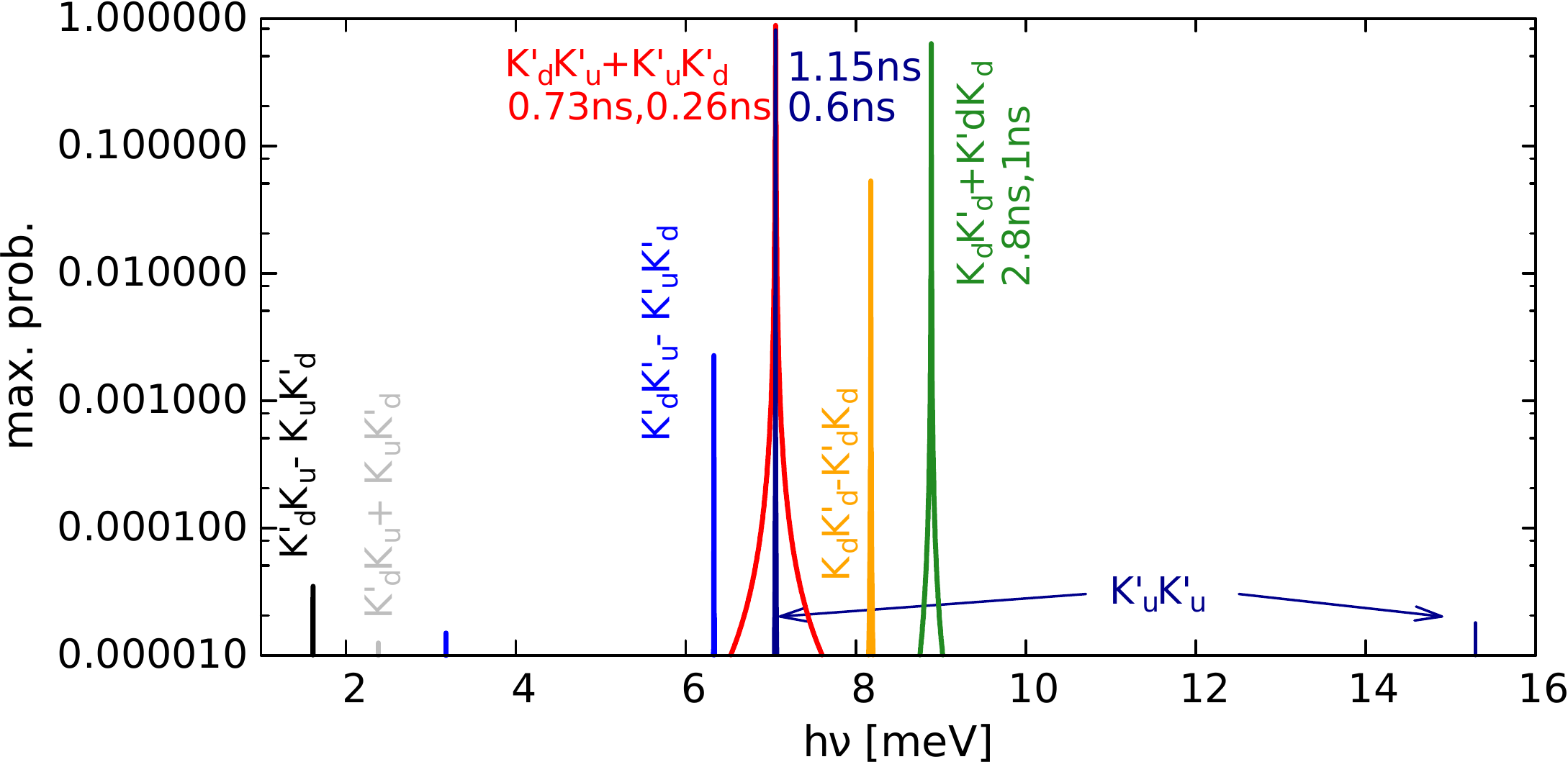} 
\caption{
Simulation for $l=18$ nm and $B=1$ T and the weekly asymmetric quantum dots of Fig. \ref{eper}(a).  The initial state
is the spin-valley polarized state with both electrons in $K'_d$ energy level.
Maximal occupation probabilities for the AC electric field applied for 3.7 ns and $F=2$ kV/cm are given by the solid lines.   The two values in ns give
the time needed to increase the occupation of the eigenstate 
above 10\% and 50\%. 
}
 \label{zczasemkoniec}
\end{figure}

\section{Double quantum dot}

\subsection{Single electron in a symmetric pair of quantum dots}
For description of the double quantum dot
we consider a generalization of the model confinement of Eq. (\ref{numer})
\begin{eqnarray}
V^A({\bf r})&=& w+\min\left\{V^l({\bf r}),V^u({\bf r})\right\},\\
V^B({\bf r})&=&-w+\min\left\{V^l({\bf r}),V^u({\bf r})\right\},
 \label{2d}
\end{eqnarray}
with
\begin{eqnarray}
V^l({\bf r})&=&-w_l \exp\left({-\frac{(x+l/2)^2+y^2}{R_p^2}}\right),\\
V^u({\bf r})&=&-w_u \exp\left({-\frac{(x-l/2)^2+y^2}{R_p^2}}\right),
 \label{2d1}
\end{eqnarray}
where $w_l$ and $w_u=w$ are the potential
depths of the quantum dot below and above the $y=0$ axis, respectively,
and $l$ is the distance between the dot centers.
This form of potential for $w_l=w_u$ allows one to obtain the single dot potential both at $l=0$ and in the limit of large $l$. For the study of the double dot the vertical
edge of the flake was extended to 46.3 nm (see Fig. \ref{d1k}). 

The energy spectrum for a symmetric pair of dots $w_l=w_u$
is displayed in Fig. \ref{d1k}(a). 
For the symmetric pair of dots the eigenstates can be labelled by the parity operator eigenvalues.  
For both the valleys the continuum Hamiltonian including the Rashba interaction \cite{Ezawa12a}
for a point-symmetric potentials 
commutes with the parity operator ${P}_4={P} \left(\begin{array}{cccc} 1 & 0 & 0 & 0 \\ 0 & -1 & 0 & 0 \\ 
0 & 0 & -1 & 0 \\ 
0 & 0 & 0 & 1  \end{array}
\right)$,
which is diagonal 
in the space of four-component wave functions 
$\phi=\left(\begin{array}{l} \phi^u_A \\  \phi^u_B \\  \phi^d_A \\  \phi^d_B\end{array}\right)$,
with the two first (last) components that correspond to the spin-up (spin-down) wave functions, the subscripts denotes the sublattice. In the definition of ${P}_4$, the scalar parity operator   ${P}$ is used that inverts the argument of a scalar 
wave function  ${P}f({\bf r})=f(-{\bf r}).$

In Fig. \ref{d1k}(a) the energy levels are
labelled by the valley index $K$ or $K'$ and the 
subscript $e$ or $o$ for even and odd ${P}_4$ eigenstates
that correspond to the eigenvalue  $+1$ or $-1$, respectively. 
 The spin of the levels is 
marked by the color of the lines. 

In Fig. \ref{d1k}(d-g) we plot the ground-state electron densities as obtained with the attomistic approach for the spin-up (d,e) and spin-down (f,g) components on sublattice A (d,f) and B (e,g) for $B=0.5$ T and $l=18$ nm. 
The ground-state is the lowest odd $P_4$-parity $K'$ spin-down energy level.
The spin-down A sublattice is dominant [Fig.\ref{d1k}(f)]
as for the single-dot ground state 
and the spin-up components are very small [Fig.\ref{d1k}(d,e)],
which results from the essential weakness of the Rashba interaction in silicene. The spin-up A and the spin-down B [Fig. \ref{d1k}(d,g)] components vanish at the origin,
which results from the spatial antisymmetry of these
components with respect to the point inversion $P$. 
The two other wave function components are even eigfunctions of the scalar parity operator $P$.

The energy spectrum of Fig. \ref{d1k} at large $l$ 
is a degenerate version of the single-dot spectrum for $l=0$.
Let us discuss the energy levels 
starting from $l=30$ nm and going to lower $l$ values. 
The splitting of the energy levels at finite  $l$ results
from activation of the tunnel coupling between the dots. 
Since the A sublattice component of the wave functions is 
dominant, one can attribute the bonding character
to: spin-down odd ${P}_4$ parity states and spin-up
even ${P}_4$ parity states. The corresponding energy levels fall
in energy when $l$ is reduced below 30 nm. 
 The remaining states: 
the spin-down even-parity and the spin-up odd-parity energy levels are antibonding and increase when $l$ is reduced
below 30 nm. 

As a consequence of the tunnel coupling, the
energy levels 
of opposite bonding-antibonding character 
change their order near $l=16$ nm -- the region
enlarged in Fig. \ref{d1k}(b). 
The energy levels which correspond to the same valley 
but opposite spin enter into an avoided crossing that 
is open by the Rashba spin-orbit coupling, which
mixes the spins of the eigenstates within the range of
the avoided crossing. 

The avoided crossing of energy levels appears at a narrow range of $l$ but
it leaves a much wider signature on the transition matrix elements. They are displayed in Fig. \ref{d1k}(c)
for the spin-down $K'_o$ ground state set as the initial state
of the transition. 
The valley and spatial parity selection
rules allow the transition only to $K'_e$ states 
[cf. Fig. \ref{d1k}(c)].
Near the avoided crossing of Fig. \ref{d1k}(b), the
spin-flipping transition increases by two orders of the magnitude [Fig. \ref{d1k}(c)]. 
The result indicates that one can arrange for very fast spin flips
using double dot potentials. 

\subsection{Single electron: asymmetric double dots}

The ideal symmetry assumed for Fig. \ref{d1k}
can hardly be achieved experimentally. For that reason we considered also  asymmetric systems. 
In Fig. \ref{wasd}(a) we plotted the energy levels
as functions of the interdot distance for 
the dot defined at the lower part of the flake $y<0$
made deeper by 6 meV ($w_l=1.03w_u$). The colors in Fig. \ref{wasd}(a)
indicate the localization of the charge density.
For $l<15$ nm the electron density is distributed 
nearly equally at both sides of the $y=0$ line
which indicates a strong tunnel coupling between the dots.
For $l>15$ nm the states exhibit a distinctly stronger localization in one of the dots. 
The avoided crossing open by the Rashba interaction discussed above is still observed  [Fig. \ref{wasd}(b)].
At the center of the avoided crossing not only the spins
are exchanged, but the charge density is equally distributed
between the dots. 

Fig. \ref{wasd}(c) shows the dipole matrix elements
for the transitions from the ground-state.
For a non-symmetric system a second spin-flipping transition
appears. For the symmetric system the  transition to the higher-energy $K'_u$ state
was forbidden by the spatial symmetry: the higher-energy  $K'$  spin-up state in the symmetric quantum dot [Fig. \ref{d1k}(a)] is $P_4$-odd, similarly 
as the ground-state. 
Moreover, for the symmetric system the transition
to the spin-down even-parity state constantly grows  
with $l$ [Fig. \ref{d1k}(c)]. 
For an asymmetric system the matrix element for the 
transition from the $K_d'$ ground-state
to the first  excited state of the same spin-valley  is 
a non-monotonic function of the interdot distance [Fig. \ref{wasd}(c)] and decreases at large $l$  since the
ground and the excited state end localized in different dots 
 and the overlap between the wave functions vanishes.

The simulation of driven transitions for
$F_{AC}=200$ V/cm is displayed in Fig. \ref{czas119}. Note that we reduced $F_{AC}$ ten times with respect to Figs. \ref{czas1} and Fig. \ref{czas2}.
The  ground state at $B=0.5$T, i.e.  
 the $K'$  spin-down oriented, 
 mostly confined in the lower dot ($y<0$), is applied for the initial condition.
The occupation of any state in the $K$ valley reaches at most $\simeq 0.2\%$ during the 3.74 ns
that is covered by the computation. 
Figure \ref{czas119}(a) was calculated at $l=18$ nm -- near the center of avoided crossing in Fig. \ref{wasd}(b).
At the frequency $h\nu$ below 14 meV a Rabi spin-flip to the $K'$ state localized in the upper dot ($K'_u(y>0)$) is observed
with the half transition time of 0.79 ns. 
At $h\nu \simeq 7$ meV transitions to as many as three states occur.
The transitions are driven to the  energy levels $K'_d(y>0)$ and $K'_u(y<0)$ that enter into the avoided crossing [Fig. \ref{wasd}(b)].
The fastest is the transition to $K'_d$ state -- that amounts in the electron charge hopping from the lower to upper dot
with conserved spin and valley. The spin-flipping transition to $K'_u(y<0)$ leaves the charge in the lower dot  and lasts a few times longer.
Anyway, the spin-flip time is of the order of the charge transition thanks to the proximity of the avoided crossing open by the Rasba interaction that mixes the spins of the states involved.
The third transition for this energy is the one to $K'_u(y>0)$ for $h\nu$ which is half the direct Rabi transition. 
This two-photon transition is enhanced by the overlap with the proximity of the transition to $K'_u(y<0)$. The electron first flips its spin passing from the ground state to $K'_u(y<0)$ and next jump to the other dot to $K'_u(y>0)$.
The peaks at lower energy in Fig. \ref{czas119} are the two-photon and three-photon transitions to the discussed states. 

Figure \ref{czas119}(b) was taken for $l=24$ nm, far from the avoided crossing of Fig. \ref{zczasemkoniec}(b).
The half-time for the charge hopping to $K'_d(y>0)$ states is only 3ps. The spin-flip occurs faster with 
the charge hopping to the other dot ($K'_u(y>0)$, 1.6ns) than the spin-flip within the dot [$K'_u(y<0)$].
The change of the order of spin flip times with and without charge hopping  is consistent with the result
of Fig. \ref{zczasemkoniec}(c) for the matrix elements of transitions to $K'_u$ states. 

As the last case for the single-electron in a double dot we consider a stronger asymmetry of the potentials
in Fig. \ref{asy11} with  $w_l=1.1w_u=0.22$ eV.
For this potential difference the avoided crossing [Fig. \ref{d1k}(b), Fig. \ref{wasd}(b)] no longer occurs [Fig.\ref{asy11}(a)].
However, the spin-flipping matrix elements [Fig. \ref{asy11}(b)] still possess a local maximum for
$l$ between 15 and 18 nm. 

The driven transitions  are displayed in Fig. \ref{czasyasy11} for $l=18$ nm (a)
and $l=24$ nm (b). Here, the  amplitude was increased back to $F_{AC}=2$ kV/cm. 
The spin-flip is about three times faster for the transition to the other dot (to $K'_u(y>0)$) 
than the intradot transition (to $K'_u(y<0)$), and the interdot distance $l$
changes the transition times by a factor of $\simeq 3$. 
For both $l$ considered in Fig. \ref{czasyasy11} we observe fast charge hopping transitions (red lines) 
with the single-photon Rabi resonance above $h\nu_R=15$ meV and a series of $n$-photon transitions
for the energies of $\nu=\nu_R/n$.

\subsection{The electron pair}

Figures \ref{eper}(a) and \ref{eper}(c) show the spectrum for the interdot distance of $l=18$ nm 
for asymmetric and symmetric double dots, respectively. 
For the electron pair the asymmetry of the double dot potential is of limited importance
for the charge distribution, since the Coulomb interaction keeps the carriers localized in separate dots,
so the spectra of Fig. \ref{eper}(a) and Fig. \ref{eper}(c) are qualitatively identical. 
The separation of electrons makes the intervalley exchange  discussed for a single dot in Fig. \ref{ep} negligible. 

For $B=0$ we observe three groups of states.
In the lowest (the highest) group including 4 levels both the electrons occupy the states of the ground-state (excited state)
 doublet $K'_d$ or $K_u$ ($K_d$ or $K'_u$) for each of the dots [Fig. \ref{widma1eb}(b)].
In the central group of 8 energy levels an electron in one dot occupies a state
of the ground-state doublet ($K'_d$ or $K_u$) and an electron in the other dot occupies the excited state ($K_d$ or $K_u$). 

If one neglects the intervalley coupling -- by the edge, and the contribution
of both spins to the wave function -- due to the weak Rashba interaction,
 the wave functions for the electron pair in separate dots
can be interpreted in terms of separable products of spin-valley ($\psi_{sv}$) and spatial wave functions
($\psi_{sp}$), i.e., $\Psi(1,2)=\psi_{sv}(1,2)\psi_{sp}(1,2)$.
Both the spatial and spin-valley parts have a definite and opposite symmetry
with respect to the electron interchange. 

For carriers localized in separate dots and four accessible spin-valley single-electron states ($K_d,K_u,K'_d,K'_u$) 
the two-electron spin-valley $\psi_{sv}$ wave functions takes one of the 16 forms
which are explicitely given  close to the energy levels in Fig. \ref{eper}(a). 
The normalization is skipped in the formulae given in the Figure, and in the products of single-electron spin-valley
terms the first term corresponds to the first electron and the other to the second,
i.e. $K_dK_d$ stands for $K_d(1)K_d(2)$, etc.
In Fig. \ref{eper}(a) one finds ten symmetric and six antisymmetric spin-valley wave functions with respect to the electron interchange.  Four of the ten symmetric functions are 
spin-valley polarized ($K_d(1)K_d(2)$ etc.). 
The twelve remaining states which are not spin-valley polarized appear in both symmetric and antisymmetric forms and the corresponding energy
levels shift in pairs as functions of the magnetic field [Fig. \ref{eper}(a)].
The spatial part of the wave function can be in the crudest approximation expressed as $\psi_{sp}(1,2)=\frac{1}{\sqrt{2}}\left(\phi_1({\bf r}_1)\phi_2({\bf r}_2)\pm \phi_2({\bf r}_1)\phi_1({\bf r}_2)\right)$,
where $\phi_1$ and $\phi_2$ are the single-electron orbitals. 
%and the sign is plus (minus) for the states symmetric (antisymmetric) with respect to the electron interchange. 
For the antisymmetric $\psi_{sv}$ the orbital part needs to be symmetric. For the symmetric spatial function both the electrons can occupy  the same e.g. the bonding single-electron orbital, hence the redshift of the antisymmetric spin-valley states with respect to the symmetric ones in Fig. \ref{eper}(a). 
The splitting of the energy levels that shift is pairs is due to the tunnel coupling between the dots and is known
from  the quantum dots in 3D materials \cite{losb}. Fig. \ref{eper}(b) provides the result for the interdot distance increased to $l=24$ nm.
Here, the splitting of the symmetric and antisymmetric states can no longer be resolved
since the tunnel coupling between the dots is already quenched. In the absence of the valley degree of freedom,
the splitting appears between symmetric and antisymmetric spin states and 
is referred to as the ''spin exchange energy''  \cite{losb}.

Figure \ref{zczasemkoniec} shows the results of the simulation
for the transitions driven at 1T from the spin-valley polarized ground-state $K'_d(1)K'_d(2)$.
The asymmetric dots of Fig. \ref{eper}(a) are taken, with the magnetic field of 1T,
 the $F_{AC}$ amplitude of 2kV/cm, and the duration of the AC pulse of 3.74 ns.
We do not observe effective ($>10\%$) transitions from the $K'_d(1)K'_d(2)$ ground state to the remaining
three-states of the lowest-energy  group of energy levels of Fig. \ref{eper}(a), since these transitions 
require both the spin and the valley flip for one of the electrons.
For the central group of energy levels relatively fast transitions are observed to $K'_d(1)K'_u(2)+K'_u(1)K'_d(2)$ 
and $K_d(1) K'_d(2)+K'_d(1) K_d(2)$ states. These spin-valley states are symmetric with respect to the electron interchange
and require either the spin or the valley flip in the ground-state $K'_d(1)K'_d(2)$ wave function.
Note, that the transitions driven by the $F_{AC}(y_1+y_2)\sin(h\nu t)$ perturbation would be strictly forbidden
between the exactly separable states of opposite symmetry of the spatial part of the wave function.

For the highest group of the energy levels one observes only the transitions to the $K'_u K'_u$ energy level.
This transition requires the spin flip of both the electrons in the ground state $K_d'K_d'$. 
The corresponding transition peak is very small at the nominally resonant energy of about $h\nu =15$ meV, 
but a high transition peak is found at half the resonant energy. The peak overlaps with the transition to $K'_d(1)K_u(2)+K'_u(1)K'_d(2)$ state.
Hence, the transition to $K'_uK'_u$ at the energy if $\simeq$ 7 meV has an indirect two-photon character and occurs by sequential flips of the two spins, one after the other.

\section{Summary and Conclusions}

We studied the spin and valley properties of electrons confined
in single and double electrostatic quantum dots defined within the silicene
using atomistic tight binding and exact diagonalization approach
for  both stationary Hamiltonian eigenstates
 and the spin-valley dynamics of the system driven by microwave or deep infrared field.

The effects of the intervalley exchange interaction and the interdot tunnel
coupling on the spectra confined in single and double quantum dots were explained. 
We determined the transition rates involving charge hopping, spin flipping and valley  switching.
We found that the valley-transition times can be changed by orders of magnitude by tuning the coupling of the confined system to the crystal edge by
the electrostatic confinement potential. With the control of the electrostatic confinement one can both produce very fast valley flips or remove the orbital-valley coupling.
The intervalley scattering due to the electron-electron interaction increases
the valley transitions times to a significant extent only when the coupling to the edge
is removed. The spin-transition times depends on the strength electrostatic confinement is weaker. Ultrafast  spin transitions in the double dots 
due to the coupling of the bonding and antibonding orbitals by the Rashba interaction were demonstrated.

\section*{Acknowledgments}
This work was supported by the National Science Centre (NCN) according to decision DEC-2016/23/B/ST3/00821
and by the Faculty of Physics and Applied Computer Science AGH UST statutory research tasks within the subsidy of
Ministry of Science and Higher Education. The calculations were performed
on PL-Grid Infrastructure.

\end{document}